\documentclass[aip,jcp,preprint,longbibliography]{revtex4-1}
\usepackage{graphicx}
\usepackage{amsmath, amssymb}% Include figure files
\usepackage{color}
\usepackage{ulem}
\def\ptl{\partial}
\def\ie{\textit{i.e.}, }
\def\eg{\textit{e.g.}, }
\def\glv{\gamma_\mathrm{LV}}
\def\gsl{\gamma_\mathrm{SL}}
\def\gsv{\gamma_\mathrm{SV}}
\def\gs0{\gamma_\mathrm{S0}}
\def\gl0{\gamma_\mathrm{L0}}
\def\gv0{\gamma_\mathrm{V0}}
\def\Wsl{W_\mathrm{SL}}
\def\Wsv{W_\mathrm{SV}}

\def\rsf{r_\mathrm{SF}}

\def\rlblk{r_\mathrm{L}^\mathrm{blk}}
\def\rvblk{r_\mathrm{V}^\mathrm{blk}}
\def\dxizdz{\mathrm{d} \xi_{z}/\mathrm{d}z}
\def\xizcl{\xi_{z}^\mathrm{cl}}
\def\xiztop{\xi_{z}^\mathrm{top}}
\def\xizbot{\xi_{z}^\mathrm{bot}}

\def\eqref#1{(\ref{#1})}

\def\bm#1{\mbox{\boldmath $#1$}}
\def\xf{x_\mathrm{f}}

\def\zf{z_\mathrm{f}}
\def\rf{r_\mathrm{f}}
\def\vtf{\vartheta_\mathrm{f}}
\def\zpf{z{'}_\mathrm{f}}
\def\rpf{r{'}_\mathrm{f}}

\def\dzf{\mathrm{d}z_\mathrm{f}}
\def\drf{\mathrm{d}r_\mathrm{f}}
\def\zc{z_\mathrm{c}}

\def\zs{z_\mathrm{s}}
\def\vts{\vartheta_\mathrm{s}}

\def\zds{z{'}_\mathrm{s}}
\def\dzs{\mathrm{d}z_\mathrm{s}}

\def\zsl{z_\mathrm{SL}}

\def\zflr{z_\mathrm{flr}}
\def\zceil{z_\mathrm{ceil}}

\def\rnf{\rho_{V}^\mathrm{f}}
\def\rns{\rho_{A}^\mathrm{s}}
\def\usl{u_\mathrm{SL}}
\def\usv{u_\mathrm{SV}}
\def\usf{u_\mathrm{sf}}
\def\plblk{p_\mathrm{L}^\mathrm{blk}}
\def\pvblk{p_\mathrm{V}^\mathrm{blk}}
\def\rc{r_\mathrm{c}}
\def\drom{\mathrm{d}}
\def\rcv{r_\mathrm{CV}}
\def\rs{R_\mathrm{s}}
\def\Rs{R_\mathrm{s}}

\def\FzdiagSL{F_{z}^\mathrm{diag(SL)}}
\def\FzdiagSV{F_{z}^\mathrm{diag(SV)}}
\newcommand{\osaka}{Department of Mechanical Engineering, Osaka University, 2-1 Yamadaoka, Suita 565-0871, Japan}
\newcommand{\tuswater}{Water Frontier Research Center (WaTUS),
Research Institute for Science \& Technology,
Tokyo University of Science,
1-3 Kagurazaka, Shinjuku-ku, Tokyo, 162-8601, Japan}
\newcommand{\osakac}{Deptartment of Mechanical Engineering, 
Osaka City University, 3-3-138 Sugimoto, Sumiyoshi, Osaka 558-8585, Japan}
\begin{document}
\begin{flushright}
{\small The following article has been submitted to The Journal of Chemical Physics.}
{\small \copyright 2021 Yasutaka Yamaguchi. This article is distributed under a Creative Commons Attribution (CC BY) License.
}
\end{flushright}
%\preprint{}

\title{Curvature dependence of the interfacial tensions around nanoscale cylinder: Young's equation still holds
}

\author{Keitaro Watanabe}
\altaffiliation{present affiliation: Department of Management of Industry and Technology, Osaka University}
\affiliation{\osaka}
\author{Hiroki Kusudo}
\email{hiroki@nnfm.mech.eng.osaka-u.ac.jp}
\affiliation{\osaka}
\author{Carlos Bistafa}%
\email{bistafa@nnfm.mech.eng.osaka-u.ac.jp}
\affiliation{\osaka}
\author{Takeshi Omori}%
\email{omori@osaka-cu.ac.jp}
\affiliation{\osakac}
\author{Yasutaka Yamaguchi}
\email{yamaguchi@mech.eng.osaka-u.ac.jp}
%\homepage{http://www-nnfm.mech.eng.osaka-u.ac.jp/~yamaguchi/}
\affiliation{\osaka}
\affiliation{\tuswater}
%
%
%\author{Laurent Joly} 
%\affiliation{\ilm} 
%
\date{\today}

\begin{abstract}
By extending the theoretical framework derived in our previous study [Y. Imaizumi et al., J. Chem. Phys. 153, 034701 (2020)], we successfully calculated the solid-liquid (SL) and solid-vapor (SV) interfacial tensions of a simple Lennard-Jones fluid around solid cylinders with nanometer-scale diameters from single equilibrium molecular dynamics (MD) systems, in which a solid cylinder was vertically immersed into a liquid pool.
The SL and SV interfacial tensions $\gsl - \gs0$ and $\gsv - \gs0$ relative to that for bare solid surface $\gs0$, respectively were obtained by simple force balance relations on fluid-containing control volumes set around the bottom and top of the solid cylinder, which are subject to the fluid stress and the force from the solid.
The theoretical contact angle calculated by Young's equation using these interfacial tensions agreed well with the apparent contact angle estimated by the analytical solution fitted to the meniscus shape, showing that Young's equation holds even for the menisci around solids with nanoscale curvature.
We have also found that the curvature effect on the contact angle was surprisingly small while it was indeed large on the local forces exerted on the solid cylinder near the contact line. 
In addition, the present results showed that the curvature dependence of the SL and SV interfacial free energies, which are the interfacial tensions, is different from that of the corresponding interfacial potential energies.
%, and a strong curvature effect should appear rather in the local force exerted on the solid cylinder around the contact line than in the contact angle.
\end{abstract}

\maketitle %\maketitle must follow title, authors, abstract and \pacs
\section{Introduction}
\label{sec:intro}
As we see cap-shaped liquid droplets on solid surfaces almost everyday, wetting behavior is one of the most common physical phenomena in human life, and is also a topic of interest in various scientific and engineering fields.~\cite{deGenne1985, Ono1960, Rowlinson1982, Schimmele2007, Drelich2019}
%
%, \textit{e.g.,} in the fabrication process of semiconductors,~\cite{Tanaka1993} 
%where the length scale of the structure has reached 
%down to several nanometers.\cite{Yamaguchi2018JJMF,Yamaguchi2019ECSTrans} 
By introducing the concept of interfacial tensions and contact angle $\theta$, wetting is usually described by Young's equation~\cite{Young1805} given by
 \begin{equation}
   \gsl-\gsv+\glv \cos\theta = 0,
   \label{eq:Young}
 \end{equation}
where $\gsl$, $\gsv$ and $\glv$ denote
solid-liquid (SL), solid-vapor (SV) and liquid-vapor (LV) interfacial 
tensions, respectively. 
Young's equation~\eqref{eq:Young} was first proposed based on the wall-tangential
force balance of interfacial tensions exerted on the contact line (CL) in 1805 -- before the establishment 
of thermodynamics~\cite{Gao2009} --, now it is often explained from a thermodynamic 
point of view instead of the mechanical force balance.~\cite{deGenne1985} 
Practically, the contact angle is used as a common measure of wettability.
% at the macroscopic scale.
%
%\par
%In addition, the concept of the precursor film was also introduced,
%by de Gennes,~\cite{deGenne1985} and available experimental observations of the precursor films were summarized in a review paper.~\cite{Popescu2012}
% % %
Various models have been proposed to capture the details of the CL, such as introducing the concept of precursor film~\cite{deGenne1985,Popescu2012}
or 
the microscopic contact angle,~\cite{white1977deviations} or  considering the effects of line tension due to the contact-line curvature in Eq.~\eqref{eq:Young}.~\cite{Boruvka1977, Marmur1997line}
However, it is difficult to experimentally validate these models mainly because measuring the interfacial tensions $\gsl$ and $\gsv$, which include the solid phase, is not trivial.~\cite{Kumikov1983,Tyson1977}
% %
\par
Wetting plays a key role especially in the nanoscale with a large surface to volume ratio, and from a microscopic point of view, \citet{Kirkwood1949} first provided the theoretical framework of surface tension based on the statistical mechanics. Recent development of molecular simulation methods including molecular dynamics (MD) and Monte Carlo (MC) advanced the microscopic understanding of the interfaces, and the calculation of the surface tension based on Bakker's equation,~\cite{Bakker1928,Ono1960,Rowlinson1982} which describes the relation between the stress integral around the liquid-vapor or liquid-gas interface and the surface tension, is used as a standard approach.~\cite{Allen1989} In addition, MD or MC studies about microscopic wetting have been conducted, ranging from simply evaluating the apparent contact angle, \eg from the average droplet shape, to quantitatively extract the SL and SV interfacial tensions through a mechanical manner and/or a thermodynamic manner.~\cite{Tang1995, Gloor2005, Ingebrigtsen2007, Das2010, Weijs2011, Seveno2013, Nijmeijer1990_theor, Nijmeijer1990_simul, Nishida2014, Imaizumi2020, Surblys2014, Yamaguchi2019, Kusudo2019, Leroy2009, Leroy2010, Leroy2015,  Kanduc2017, Kanduc2017a, Surblys2018, Bistafa2021,  Grzelak2008, Lau2015, Kumar2014, Ardham2015, Jiang2017,  Ravipati2018, Omori2019, Bey2020}
For the mechanical approach -- called the mechanical route --, Bakker's equation was extended to describe the connection between the stress integral through the SL or SV interface and the corresponding interfacial tension (See Appendix~\ref{appsec:extbakker_cylinder}).~\cite{Nijmeijer1990_theor, Nijmeijer1990_simul, Nishida2014, Imaizumi2020, Surblys2014, Yamaguchi2019, Kusudo2019} On the other hand for the latter approach -- called the thermodynamic route --, the SL and SV interfacial tensions were interpreted as the interfacial free energy per interfacial area. For instance, by the thermodynamic integration method, the SL interfacial energy was evaluated as the free energy difference from a reference system, in which the SL interface was substituted by bare solid and liquid surfaces  quasi-statically under constant number of particles $N$, temperature $T$ and pressure $p$ ($NpT$-ensemble) or volume $V$ ($NVT$-ensemble) condition.~\cite{Surblys2014, Yamaguchi2019, Kusudo2019, Leroy2009, Leroy2010, Leroy2015,  Kanduc2017, Kanduc2017a, Surblys2018, Bistafa2021}
These studies indicated that the apparent contact angle of the meniscus or droplet obtained in the 
simulations agreed well with the one predicted by Young's equation~\eqref{eq:Young} in case the solid surfaces are flat and smooth so that the CL pinning may not be induced.~\cite{Yamaguchi2019, Kusudo2019}
\par
Considering the potential applications of nano-wetting, \eg a flow in a confined space such as a nanofoam~\cite{Rode1999} or a carbon nanotube,~\cite{Falk2014}
the solid surfaces can have a nanoscale radius of curvature, and the interfacial tensions should depend on the curvature. 
Regarding the water wetting on carbon nanotubes as a solid with a nanoscale curvature, unique wetting behavior~\cite{Homma2013} and a strong diameter dependence of the capillary force were experimentally reported.~\cite{Imadate2018}
For the liquid-vapor interface, 
\citet{Tolman1949} first formulated the size effect of droplet surface tension with a lengthscale called  the ``Tolman length,"~\cite{Blokhuis2006,Tumram2017,Elliott2021} and MD or MC simulations have been carried out as well.~\cite{Yaguchi2010,Das2011b,Lau2015,Cheng2018,Rehner2018,Gao2021}
Indeed, the LV interfacial tension can be extracted using a strict definition of the interface position, \eg based on the force and momentum balances,~\cite{Yaguchi2010} and the difference between the pressures inside and outside the droplet based on the Young-Laplace equation. On the other hand, the calculation of the SL and SV interfacial tension on a curved solid surface is not trivial.~\cite{Cheng2018,MonterodeHijes2019}
For instance, via thermodynamic routes, if we suppose calculation systems for the thermodynamic integration to calculate the SL interfacial tension on a cylindrical solid surface with a nanoscale radius, then the three interface areas, \ie the radii of the target solid-fluid interface, the reference bare solid and bare liquid interfaces should be all different, and this difference would become critical for the evaluation of the desired interfacial tension when the cylinder radius is comparable to the radius differences.
%
%would is of course negligible in the macroscopic scale, but...
% About the calculation methods. Problem with TI regarding curved interface.
% In the nanoscale, the surface area of liquid-vapor interface changes during the stripping process.
%
%\par
Another possibility is via mechanical routes, and it is technically possible to calculate the stress distribution in the cylindrical or spherical coordinates, although the calculation cost significantly incrases to obtain the distribution, and indeed, precise calculation in these coordinate systems is not implemented into the MD packages such as  LAMMPS~\cite{LAMMPS1995} or GROMACS~\cite{GROMACS2018}, and the implementation into in-house codes is also rather complicated.~\cite{Thompson1984,Yaguchi2010} 
\par
Going back to the relation between the SL or SV interfacial tension and the fluid stress in the interface via the mechanical routes, what we need is not the stress distribution but the stress integral. Considering this feature, in our previous study,~\cite{Imaizumi2020} we provided a theoretical framework to extract the SL and SV interfacial tensions from a single MD simulation by using the local forces and the local interaction potential exerted on a quasi-two-dimensional (2D) flat and smooth solid plate immersed into a liquid pool of a simple liquid, called the Wilhelmy plate, and verified through the comparison between the MD results and the interfacial works of adhesion obtained by the thermodynamic integration (TI). This modified Wilhelmy method is advantageous because it does not require  computationally demanding calculations such as the local stress distributions and the thermodynamic integration which needs averaging at each discrete states along the integration path.
\par
In this study, we extracted the SL and SV  interfacial tensions of a simple Lennard-Jones fluid around solid cylinders with nanometer-scale radii by applying the 
modified Wilhelmy equations derived in our previous study~\cite{Imaizumi2020} to investigate the curvature effect on the SL and SV interfacial tensions.
From the results, we also examined whether Young's equation holds even for menisci around solids with nanoscale curvature. Finally, we discuss the difference in the curvature dependence between the SL and SV interfacial tensions (free energy) and the SL and SV interaction potential energies (part of internal energy).
\section{Method}
\label{sec:method}
\subsection{MD Simulation}
\begin{figure}
  \begin{center}
    \includegraphics[width=1.0\linewidth]{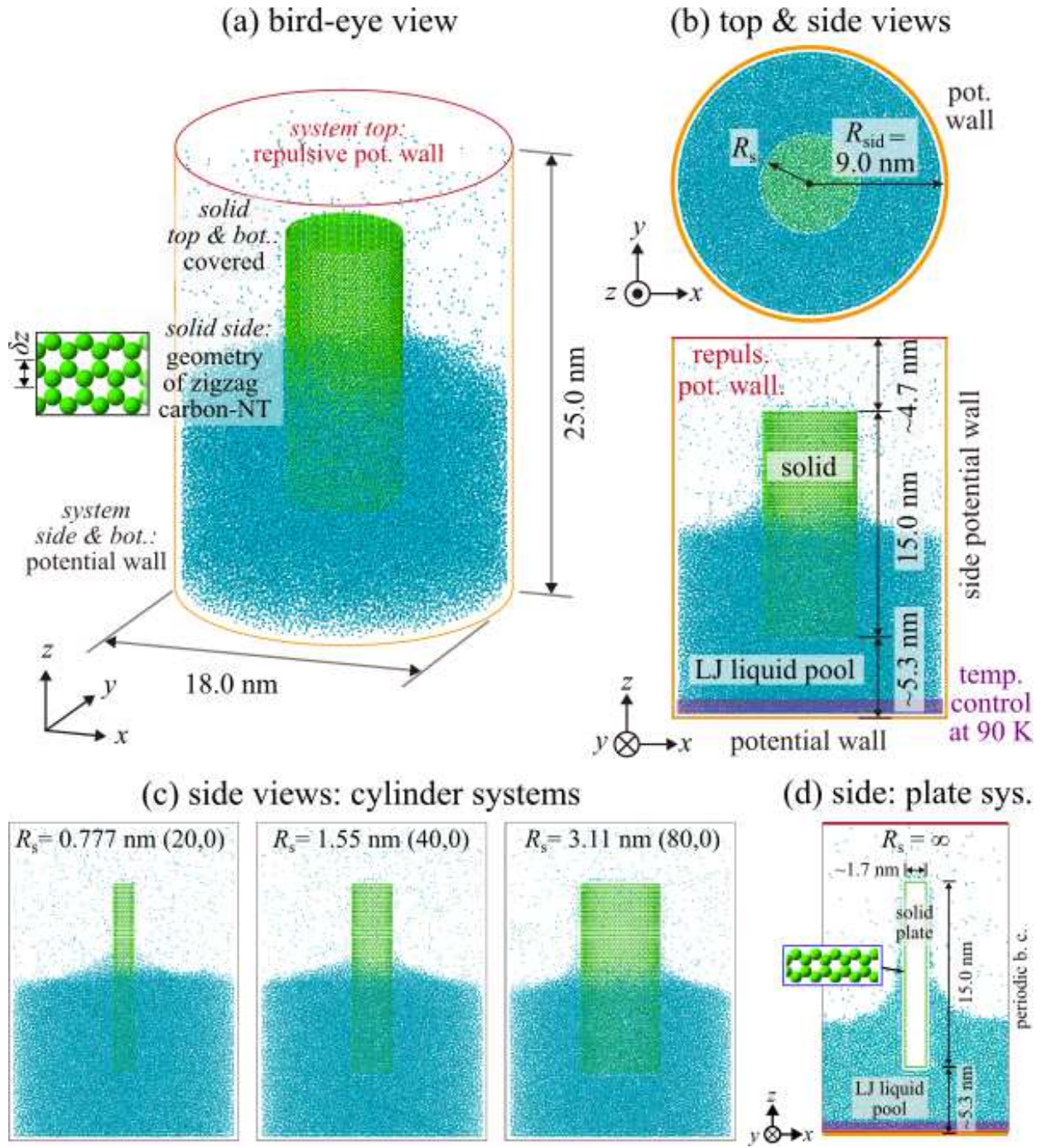}
  \end{center} 
  \caption{\label{Fig:system}
(a) A bird-eye view and (b) top and (c) side views of 
equilibrium molecular dynamics (MD) simulation systems of hollow solid cylinders dipped into a liquid pool of a simple Lennard-Jones (LJ) fluid. (d) Side view of the system with a hollow solid plate [see Ref.~\citenum{Imaizumi2020} for details].
}
\end{figure}
In this study, we employed equilibrium MD simulation systems of a quasi-axisymmetric meniscus on a hollow cylinder dipped into a liquid pool of a simple fluid as shown in Fig.~\ref{Fig:system}. Except the boundary condition
in the lateral directions, the basic setup is similar to our 
previous study of quasi-2D meniscus formed on a hollow 
rectangular solid plate.~\cite{Imaizumi2020}
%
%We call this system the `Wilhelmy MD system' hereafter.
%
Generic particles interacting through a LJ potential 
were adopted as the fluid particles. The 12-6 LJ potential 
given by
\begin{equation}
% \nonumber
   \Phi^\mathrm{LJ}(r_{ij}) = 
%   \Theta(
%   \rc
%   - r_{ij}) 
%   \\&\times& 
  4\epsilon \left[
    \left(\frac{\sigma}{r_{ij}}\right)^{12} 
    -
    \left(\frac{\sigma}{r_{ij}}\right)^{6} 
    +
    c_{2}^\mathrm{LJ}\left(\frac{r_{ij}}{\rc}\right)^2 
    +
    c_{0}^\mathrm{LJ}
  \right],
  \label{eq:LJ}
\end{equation}
was used for the interaction between fluid particles, 
where $r_{ij}$ is the distance between the particles $i$ at position 
$\bm{r}_{i}$ and $j$ at $\bm{r}_{j}$, while $\epsilon$ and $\sigma$ denote the LJ energy and length parameters, respectively. This LJ interaction was truncated at a cut-off distance of $\rc=3.5\sigma$ and quadratic functions were added so that the potential and interaction force smoothly vanished at $\rc$.
%with the Heaviside step function $\Theta$. 
The constant values of
$c_{2}^\mathrm{LJ}$ and $c_{0}^\mathrm{LJ}$
were given in our previous study.~\cite{Nishida2014}
Hereafter, fluid and solid particles are denoted by `f' and `s', respectively and corresponding combinations are indicated by subscripts.
\par
Three solid cylinders in contact with the fluid were prepared by using the geometrical configuration of carbon nanotubes (CNTs) with
their chiral indices of (20,0), (40,0) and (80,0), where the solid particles were fixed on the coordinate with the positions of hexagonal periodic structure with an inter-particle distance $r_\mathrm{ss}$ of 0.141~nm. Note that the solid surface is considered to be smooth with $r_\mathrm{ss}$ much smaller than $\sigma_\mathrm{ff}$ and $\sigma_\mathrm{sf}$, and pinning is not induced on this surface.
The corresponding radii $\Rs$ of the cylinder 
are 0.777, 1.55 and 3.11~nm, respectively.
The central axis of the cylinders is set on the $z$-axis, \ie the zigzag edge of the honeycomb structure was set parallel to the $xy$-plane. The top and bottom parts of the cylinders are covered by locating additional solid particles
as the lid to prevent fluid particles from entering into the cylinder. Note that the structure of these lids does 
not have direct effect on the simulation results as indicated
in our previous study.~\cite{Imaizumi2020}
%with locating solid 
%particles at the edge to match the hexagonal periodicity.
%The right and left faces were set at $x=\pm x_\mathrm{s}$ parallel
%to the $yz$-plane, and the top and bottom faces were parallel 
%to the $xy$-plane. 
%Note that the distance between the left and 
%right faces $2x_\mathrm{s}\approx 1.7$~nm was larger than the 
%cutoff distance $\rc$.
%
\par
The solid-fluid (SF) interaction, which denotes SL or SV interaction, was also expressed by the LJ potential in Eq.~\eqref{eq:LJ}, where the length parameter $\sigma_\mathrm{sf}$ was given by the Lorentz mixing rule, while the energy parameter $\epsilon_\mathrm{sf}$ was changed in a parametric manner by multiplying a SF interaction coefficient $\eta$ to the base value $\epsilon^{0}_\mathrm{sf}=\sqrt{\epsilon_\mathrm{ff}\epsilon_\mathrm{ss}}$ as
\begin{equation}
\label{eq:def_eta}
\epsilon_\mathrm{sf} = \eta \epsilon^{0}_\mathrm{sf}.
\end{equation}
%
%where the base value  was given by the Berthelot mixing rule as $\epsilon^{0}_\mathrm{sf} $. 
This parameter $\eta$ expressed the wettability, \ie $\eta$ and the contact angle of a hemi-cylindrically shaped equilibrium droplet on a homogeneous flat solid surface had a one-to-one correspondence~\cite{Nishida2014, Yamaguchi2019, Kusudo2019}, and 
we set the parameter $\eta$ between 0.03 and 0.15 so that 
the corresponding cosine of the contact angle $\cos \theta$ 
is from $-0.9$ to $0.9$. The definition of the 
contact angle is described later in Sec.~\ref{sec:resdis}.
Note that due to the fact that the solid-solid inter-particle distance 
$r_\mathrm{ss}$ shown in  Table~\ref{tab:table1}
were relatively small compared to  the LJ length parameters 
$\sigma_\mathrm{ff}$ and $\sigma_\mathrm{fs}$, the surface is considered 
to be very smooth, and the wall-tangential force from the solid on the 
fluid, which induces pinning of the CL, is 
negligible.~\cite{Yamaguchi2019, Kusudo2019}
\par
We set a horizontal 
potential wall on the bottom (floor) of the calculation cell 
fixed at $z=\zflr$ about 5.3~nm
%\wata{checked! Strictly 5.2875~nm} 
below the bottom of the solid plate, which interacted only with 
the fluid particles with a one-dimensional potential field $\Phi_\mathrm{flr}^\mathrm{1D}$
as the function of the distance from the wall given by
\begin{gather}
  \label{eq:potentialbath}
  \Phi_\mathrm{flr}^\mathrm{1D}(z'_{i})=
%  \Theta(r^\mathrm{c}_\mathrm{sf} - z{''}_i) 
%  \cdot 
  4\pi \rho_{n} \epsilon^\mathrm{flr}_\mathrm{sf}
  \sigma_\mathrm{sf}^{2}
  \left [ 
    \frac{1}{5} \left(
      \frac{\sigma_\mathrm{sf} }{z'_{i} }
    \right)^{10}
    \!\!\!\! - 
    \frac{1}{2} \left(
      \frac{\sigma_\mathrm{sf} }{z'_{i} }
    \right)^{4}
      +
      c_{2}^\mathrm{flr}
      \left(\frac{z'_{i}}{\zc^\mathrm{flr}}\right)^2 
      +
      c_{1}^\mathrm{flr}
      \left(\frac{z'_{i}}{\zc^\mathrm{flr}}\right)
    +
    c_{0}^\mathrm{flr}
 \right],
 \\
 z'_{i}\equiv z_{i} - z_\mathrm{flr}.
\end{gather}
where $z_{i}$ is the $z$-position of fluid particle $i$, and 
$\epsilon^\mathrm{flr}_\mathrm{sf}$ is set at 
$0.09\epsilon^{0}_\mathrm{sf}$. 
This potential wall mimicked a mean potential field created 
by a single layer of solid particles with a uniform area 
number density $\rho_{n}$. 
Similar to Eq.~\eqref{eq:LJ}, this potential field 
in Eq.~\eqref{eq:potentialbath} was 
truncated at a cut-off distance of 
$\zc^\mathrm{flr}=3.5 \sigma_\mathrm{sf}$ and a quadratic function was added so that the potential and interaction 
force smoothly vanished at $\zc^\mathrm{flr}$. 
As shown in Fig.~\ref{Fig:system}, fluid particles 
were rather strongly attracted on this plane because this 
roughly corresponded to a solid wall showing complete wetting.
With this setup, the liquid pool was stably kept even 
when the liquid pressure is low with a highly wettable solid plate, 
and sufficient liquid bulk region was formed between this wall 
and the bottom of the cylinder.
Furthermore, we set another horizontal 
potential wall on the top (ceiling) of the calculation 
cell fixed at $z=\zceil$ about 4.7~nm % \wata{Checked! If the CNT is 15 nm, this distance is obtained. Strictly speaking, the length of the CNT is 14.8755 nm due to the periodicity of the structure, so the exact length of $\zceil$ is 25-14.8755-5.2875 = 4.837 nm.} 
above the top of the 
solid plate exerting a repulsive potential field $\Phi_\mathrm{ceil}^\mathrm{1D}$ 
on the fluid particles given by
\begin{gather}
  \label{eq:potentialbath_top}
  \Phi_\mathrm{ceil}^\mathrm{1D}(z''_{i})=
  4\pi \rho_{n} \epsilon^\mathrm{ceil}_\mathrm{sf}
  \sigma_\mathrm{sf}^{2}
  \left [ 
    \frac{1}{5} \left(
      \frac{\sigma_\mathrm{sf} }{z''_{i} }
    \right)^{10}
%    \!\!\!\! - 
%    \frac{1}{2} \left(
%      \frac{\sigma_\mathrm{sf} }{z''_{i} }
%    \right)^{4}
      +
      c_{2}^\mathrm{ceil}
      \left(\frac{z''_{i}}{\zc^\mathrm{ceil}}\right)^2 
    +
      c_{1}^\mathrm{ceil}
      \left(\frac{z''_{i}}{\zc^\mathrm{ceil}}\right)
    +
    c_{0}^\mathrm{ceil}
 \right],
 \\
 z''_{i}\equiv \zceil - z_{i},
\end{gather}
where $\epsilon^\mathrm{ceil}_\mathrm{sf}$ is equal to $\epsilon^\mathrm{flr}_\mathrm{sf}$
%is also 
%empirically set at $0.09\epsilon^{0}_\mathrm{sf}$. 
%\wata{In Eq. 6, the part of $\epsilon^{0}_\mathrm{sf}$ is defined by $\epsilon^{0}_\mathrm{sf} * 0.09$. }
while %where 
a cut-off distance of 
$\zc^\mathrm{ceil}=
\sigma_\mathrm{sf}$ is set to
express a repulsive potential wall.
\par
In addition to these bottom and top  potential walls, 
we also set another cylindrical side potential wall
with its axis on the $z$-axis and with a radius 
$R_\mathrm{sid}$, which exerts  a one-dimensional 
potential field on the fluid particles 
$\Phi_\mathrm{sid}^\mathrm{1D}(r^{xy}_{i})$
as the function of the horizontal distance from 
the wall given by% \red{(check potential form)}
\begin{gather}
  \label{eq:potentialbath_side}
  \Phi_\mathrm{sid}^\mathrm{1D}(r^{xy}_{i})=
%  \Theta(r^\mathrm{c}_\mathrm{sf} - z{''}_i) 
%  \cdot 
  4\pi \rho_{n} \epsilon^\mathrm{sid}_\mathrm{sf}
  \sigma_\mathrm{sf}^{2}
  \left [ 
    \frac{1}{5} \left(
      \frac{\sigma_\mathrm{sf} }{r^{xy}_{i}}
    \right)^{10}
    \!\!\!\! - 
    \frac{1}{2} \left(
      \frac{\sigma_\mathrm{sf} }{r^{xy}_{i}}
    \right)^{4}
      +
      c_{2}^\mathrm{flr}
      \left(\frac{r^{xy}_{i}}{r_\mathrm{c}^\mathrm{sid}}\right)^2 
      +
      c_{1}^\mathrm{flr}
      \left(\frac{r^{xy}_{i}}{r_\mathrm{c}^\mathrm{sid}}\right)
    +
    c_{0}^\mathrm{sid}
 \right],
 \\
 r^{xy}_{i}\equiv R_\mathrm{sid} - \sqrt{x_{i}^{2}+y_{i}^{2}},
\end{gather}
%
%\wata{checked!}
where $ r^{xy}_{i}$ is the lateral distance of fluid particle $i$
from the side potential wall. Note that this side wall was adopted 
to achieve a quasi-axisymmetric 2D-meniscus instead of applying
the periodic boundary condition in the horizontal $x$- and 
$y$-directions as in our previous study with a solid plate.
The parameter $\epsilon^\mathrm{sid}_\mathrm{sf}$ 
%\wata{$\epsilon^\mathrm{sid}_\mathrm{sf} = 0.0975 * \epsilon^{0}_\mathrm{sf}=0.192 * 10^{-21}$} 
was set at 0.0975$\epsilon^{0}_\mathrm{sf}$ so that the resulting 
contact angle at the side wall may be roughly 90 degrees.% \wata{checked!}.
\par
The temperature of the system was maintained at a constant temperature 
of $T_\mathrm{c}$ at 90~K, which is above the triple point 
temperature~\cite{Mastny2007}, by applying velocity rescaling to the 
fluid particles within 0.8~nm from the floor wall regarding the 
velocity components in the $x$- and $y$-directions. Note that this 
region was sufficiently away from the bottom of the solid plate and 
no direct thermostating was imposed on the region near the solid plate, so 
that this temperature control had no effects on the present results. 
\par
With this setting, an axisymmetric quasi-2D LJ liquid with a meniscus-shaped LV interface and the CL parallel to the $xy$-plane was formed as an equilibrium state as exemplified in Fig.~\ref{Fig:system}, 
where a liquid bulk with an isotropic density distribution existed above the bottom wall by choosing a proper number of fluid particles $N_\mathrm{f}$ as shown in Fig.~\ref{Fig:distribtution}. 
We checked that the temperature was constant in the whole system after 
the equilibration run described below.
%Note also that in the present quasi-2D systems, effects of the CL curvature 
%can be neglected.~\cite{Boruvka1977, Marmur1997line, Ingebrigtsen2007, Leroy2010, Weijs2011, Nishida2014, Yamaguchi2019,Kusudo2019} 
%
The velocity Verlet method was applied for the integration of the Newtonian equation of motion with a time increment of 5~fs for all systems. The simulation parameters are summarized in Table~\ref{tab:table1} with the corresponding non-dimensional ones, which are normalized by the 
corresponding standard values based on $\epsilon_\mathrm{ff}$,
$\sigma_\mathrm{ff}$ and $m_\mathrm{f}$. 
%Note that the values for another system we employed in Fig.~\ref{fig:SLpin} are also shown, whose details are described in Sec.~\ref{subsec:pinninglimit}.
%
\par
The physical properties of each equilibrium system with various 
$\eta$ values were calculated as the time average of 20~ns
which followed an equilibration run of more than 10~ns.
\begin{table*}[!t]
\caption{\label{tab:table1} 
Simulation parameters and their corresponding non-dimensional values.
}
%\begin{tabular*}{100mm}{@{\extracolsep{\fill}}|c|c|c|} \hline\hline
\begin{ruledtabular}
\begin{tabular}{cccc}
property  & value & unit & non-dim. value
\\ \hline
$\sigma_\mathrm{ff}$ & 0.340 & nm & 1
\\
$\sigma_\mathrm{sf}$ &  0.357 & nm & 1.05
\\
$r_\mathrm{ss}$ & 0.141 & nm & 0.415
\\
$\epsilon_\mathrm{ff}$ & $1.67 \times 10^{-21}$ & J & 1
\\
$\epsilon^{0}_\mathrm{sf}$
& $1.96\times 10^{-21}$ & J & 1.18
\\
$\epsilon^\mathrm{flr}_\mathrm{sf}$
& $0.176\times 10^{-21} $ & J & 0.106
\\
$\epsilon^\mathrm{ceil}_\mathrm{sf}$
& $0.176\times 10^{-21} $ & J & 0.106
\\
$\epsilon^\mathrm{sid}_\mathrm{sf}$
& $0.192\times 10^{-21} $ & J & 0.115
\\
$\epsilon_\mathrm{sf}$ & 
$\eta \times \epsilon^{0}_\mathrm{sf}$
\\
$\eta$ &
0.03 -- 0.15 & - & -
\\
$m_\mathrm{f}$ & $6.64 \times 10^{-26}$ & kg & 1
%\\
%$m_\mathrm{s}$ & - & - & -
\\
$\Rs$ & 0.777 -- 3.11 & nm & 2.29 -- 9.15
%\\
%$m_\mathrm{s}$ & - & - & -
\\
$T_\mathrm{c}$ & 90  & K & 0.703
\\
$N_\mathrm{f}$ & 53778 - 60834  & - & -
\end{tabular}
\end{ruledtabular}
\end{table*}
\section{Results and discussion
\label{sec:resdis}}
\subsection{Apparent contact angle}
\label{CA_force}
\begin{figure}
\centering
\includegraphics[width=1.0\linewidth]
{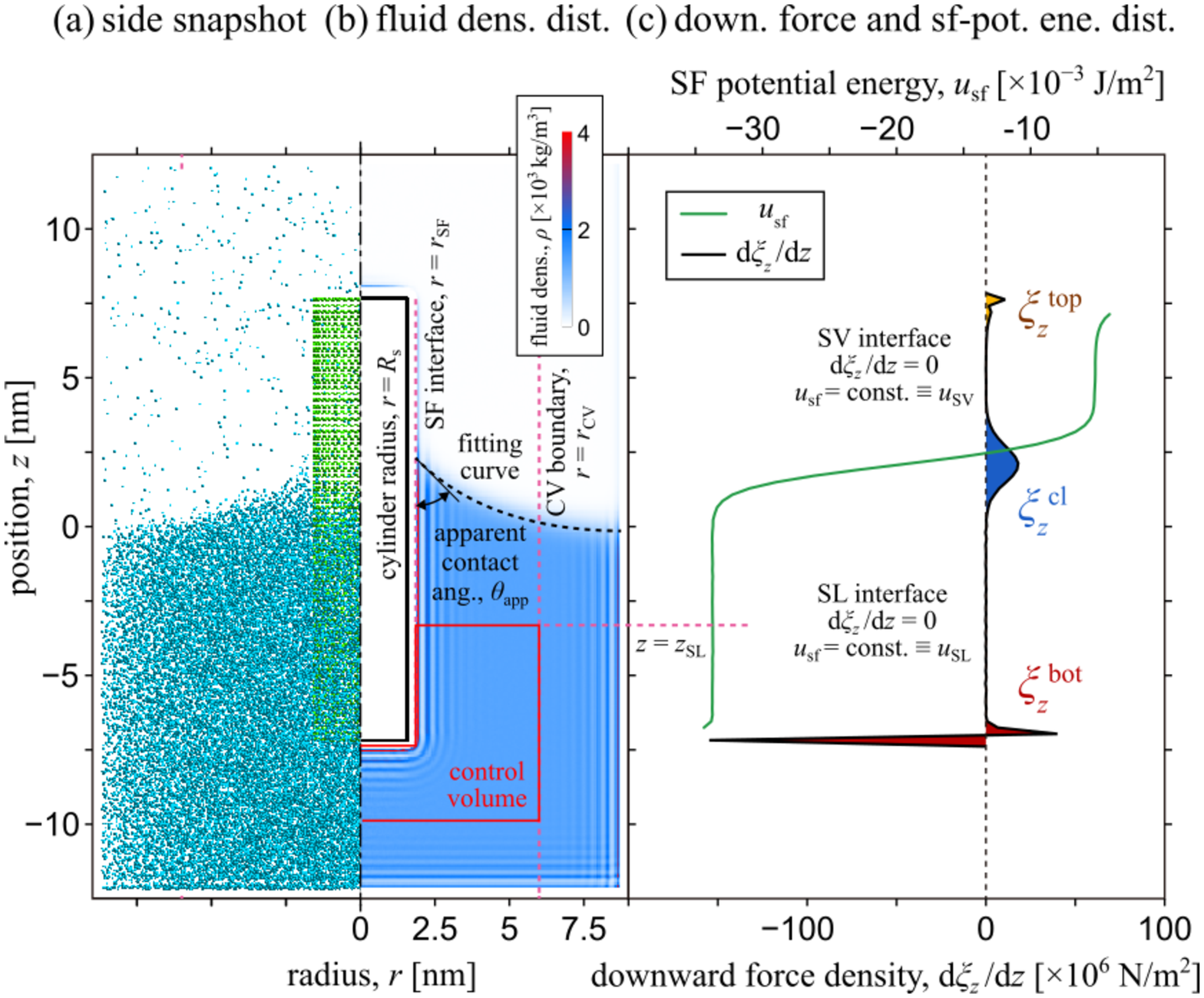}
 \caption{
(a) Half side snapshot, 
(b) distribution of the time-averaged fluid density, 
and (c) distributions of the time-averaged downward 
force density acting on the solid plate and solid-fluid (SF)
potential energy density per solid surface area for the system with the solid radius $\Rs=1.55$~nm and a
SF interaction parameter $\eta=0.15$.} 
\label{Fig:distribtution}
\end{figure}
Similar to our previous study,~\cite{Imaizumi2020} 
we calculated the distribution of force exerted from the fluid 
on the solid particles by dividing the system into equal-sized  
bins in the $z$-direction, where a bin height of 
$\delta z = $ 0.2115~nm was used, considering the periodicity 
of the CNT structure.  
We defined the average force density $\dxizdz$
as the time-averaged total downward (in $-z$-direction) force 
from the fluid on the solid particles in each bin divided by 
the solid bin area $2\pi \Rs \delta z$. 
Except at the top and bottom of the solid plate, 
$\dxizdz$ corresponds to the downward force per 
surface area. We also calculated the average SF potential 
energy per area $\usf$, which was obtained 
by the same procedure but substituting the downward 
force by the SF potential energy.
\par
Figure~\ref{Fig:distribtution} shows a half side-snapshot and 
the distribution of time-averaged fluid density $\rho$ 
around the solid cylinder for the system with solid-fluid 
interaction parameter $\eta=0.15$. The time-averaged distributions of the downward force acting on the solid plate $\dxizdz$ and the SF potential energy $\usf$ are also displayed in the right panel. We briefly summarize two essential features below,
which are qualitatively the same as in our previous study.~\cite{Imaizumi2020} 
1) Multi-layered  adsorption layers were formed around the solid 
cylinder
and the bottom and side potential walls, and liquid bulk with a homogeneous density is observed away from the cylinder, potential 
walls and the LV interface.
2) 
The total downward 
force as the integral of $\dxizdz$ can be clearly separated 
into three local parts, \ie $\xiztop$ 
around the top, $\xizcl$ around the contact line, 
and $\xizbot$ around the bottom. As indicated in 
Fig.~\ref{Fig:distribtution}~(c), $\xiztop$ and $\xizcl$ are positive, 
\ie downward forces, and 
$\xizbot$ is negative, \ie an upward force.
Note that the distributions of $\dxizdz$ and $\usf$ 
around the top and bottom had less physical meaning because 
they included the top and bottom solid lids in the bin, and these parts for $\usf$ are not displayed in the figure.
However, the 
local integral of $\dxizdz$ indeed gave the physical information 
about the force around the top and bottom parts. 
Also, note that $\xi_{z}$ has the same dimension as the surface 
tension of force per length.
\par
As exemplified in the density distribution in Fig.~\ref{Fig:distribtution}~(b), 
%From the density distribution
we evaluated the contact angles for the plate and 
cylinder systems with different solid-fluid interaction 
parameter to examine the curvature effects.
For the plate system, we followed the same procedure used to 
determine the apparent contact angle as our previous study:~\cite{Imaizumi2020}
the LV interface was defined as the least-squares fitting 
circle on the density contour of $\rho=$400~kg/m$^{3}$ 
at the LV interface at height $z(x)$ excluding 
the region in the adsorption layers near the 
solid,~\cite{Nishida2014,Yamaguchi2019,Kusudo2019,Imaizumi2020}
considering that with a constant liquid-vapor interfacial tension $\glv$ the static force balance is written as
\begin{equation}
 	\label{eq:Y-L_plate}
 	\frac{\drom}{\drom x}\left[\sin{\psi(x)}\right]
 	=\frac{\pvblk - \plblk}{\glv}, \quad
 	\tan{\psi(x)}=\frac{\drom z(x)}{\drom x},
\end{equation}
where $\psi(x)$, $\pvblk$ and $\plblk$ denote the angle from the $x$-direction and bulk pressure values in the vapor and liquid bulk, respectively. The solution of 
Eq.~\eqref{eq:Y-L_plate} results in a constant 
curvature with constant $\glv$, $\pvblk$ and $\plblk$.
Then, by setting the solid-fluid (SF) interface position $r=r_\mathrm{SF}$ as the limit that the fluid could reach, which can be easily estimated from the density 
distribution, 
we defined the apparent contact angle $\theta_\mathrm{app}$ by 
the angle between the extrapolation of the cylindrical 
LV-interface and SF-interface plane at $r=r_\mathrm{SF}$. This definition provided a 
mechanical description consistent with Young's 
equation.~\cite{Yamaguchi2019}
\par
Similarly, the contact angles on the cylinders were 
evaluated using the analytical formula of the macroscopic 
meniscus shape. For an axisymmetric equilibrium meniscus 
around a $z$-centered cylinder with neglecting gravity, 
it follows for the meniscus height $z(r)$ given as a 
unique function of the radial position 
$r\equiv\sqrt{x^{2}+y^{2}}$ that
\begin{equation}
 	\label{eq:Y-L_cylinder}
 	\frac{1}{r}\frac{\drom}{\drom r}\left[r\sin{\psi(r)}\right]
 	=\frac{\pvblk - \plblk}{\glv}, \quad
 	\tan{\psi(r)}=\frac{\drom z(r)}{\drom r},
\end{equation}
where $\psi(r)$ denote the angle from the $r$-direction.
%\par
We evaluated the three constant values 
in this differential equation~\eqref{eq:Y-L_cylinder} 
from MD simulations: 
$\glv$ was obtained from a MD system with planer LV 
interfaces by a standard mechanical process,~\cite{Surblys2014}
whereas $\pvblk$ and $\plblk$ were 
evaluated as the force per area on the top and bottom 
potential walls of the present cylinder system, 
respectively, both excluding the region near the side wall. 
Thus, by fitting the density contour of $\rho=$400~kg/m$^{3}$ 
at the LV interface excluding the region in the adsorption 
layers formed near the solid surface and also excluding 
that near the side potential wall, 
a numerical solution of the 2nd-order ordinary 
differential equation~\eqref{eq:Y-L_cylinder} can be 
obtained for each system with different cylinder radius 
$\Rs$ and solid-fluid interaction parameter $\eta$.
As shown by the dotted black line in Fig.~\ref{Fig:distribtution}, 
the meniscus shape is well reproduced in this system.
We determined the contact angle as the angle
between the extrapolated solution of the 
meniscus shape and the SF-interface position 
in the $rz$-plane including the liquid side.
We assume a concave control volume (CV) shown 
in red color in Fig.~\ref{Fig:distribtution}~(b) around 
the bottom of the cylinder to calculate the 
SL interfacial tension below. The bottom face
and the side cylindrical face of the CV are 
in the liquid bulk, where the former is sufficiently 
away from both the bottom of the system and 
bottom of the cylinder, whereas the latter 
at $r=r_\mathrm{CV}$ is away from the side 
boundary. Analogously, the top face is at the height $z=z_\mathrm{SL}$ sufficiently away from both the bottom of the cylinder and the contact line.
On the other hand, the concave faces are set 
at the solid-liquid boundary with its inner
side face at $r=r_\mathrm{SF}$.
\par
\begin{figure}
\centering
\includegraphics[width=1.0\linewidth]
{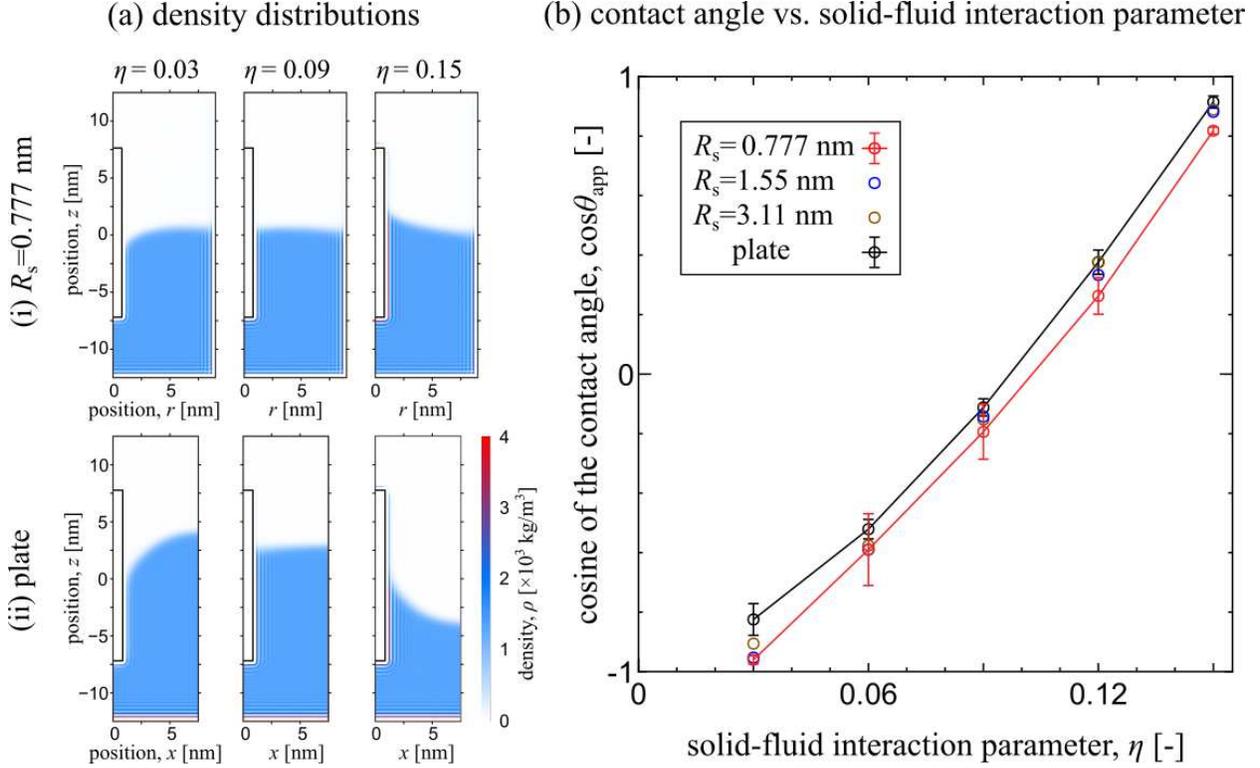}
 \caption{
(a) Density distributions around the (i) solid cylinder of 
radius $\Rs= 0.777$~nm and (ii) plate, and 
(b) relation between the cosine of contact angle and 
solid-fluid interaction parameter for different cylinder radii 
$\Rs$ and plate.
}
\label{Fig:contact_angle}
\end{figure}
The density distributions around the solid cylinder 
with the smallest radius $\Rs=0.777$~nm
and the plate, and the relation between the SF 
interaction coefficient $\eta$ and cosine of the 
contact angle $\cos \theta_\mathrm{app}$ are shown in 
Fig.~\ref{Fig:contact_angle}.
For the latter, we displayed the error bars 
and guide lines only for the plate and
for the cylinder systems with $\Rs = 0.777$~nm, for better visualization: the error bars for 
the systems with other radii were comparable to those for $\Rs = 0.777$~nm. 
As seen in Fig.~\ref{Fig:contact_angle}~(a), the apparent 
meniscus shapes of the cylinder and plate are different,
indicating that different force balances described 
by Eqs.~\eqref{eq:Y-L_plate} and \eqref{eq:Y-L_cylinder}, 
should be adopted to properly evaluate the contact angle
from the meniscus shape.
With the increase of $\eta$, the solid became more wettable,
\ie $\cos \theta_\mathrm{app}$ increased, and the cylinder with the smallest 
Radius $\Rs=0.777$~nm were less wettable, \ie had smaller 
$\cos \theta_\mathrm{app}$ than the plate for all $\eta$ values tested.
%Note also that the error bars
%
However; we should stress that the radius dependence of 
the contact angle seen in Fig.~\ref{Fig:contact_angle}~(b) 
was unexpectedly small even with the smallest cylinder with 
its radius $\Rs$ that is comparable to $\sigma_\mathrm{ff}$ or $\sigma_\mathrm{sf}$.
We discuss the reason in the following with the comparison between the apparent contact angle $\theta_\mathrm{app}$ and the contact angle predicted by Young's equation~\eqref{eq:Young} using the interfacial tensions obtained by the local forces. %About total force and Wilhelmy equation. 
\subsection{Curvature dependence of  the force around the contact line and the solid-liquid and solid-vapor interfacial tensions}
\begin{figure}
\centering
\includegraphics[width=0.8\linewidth]
{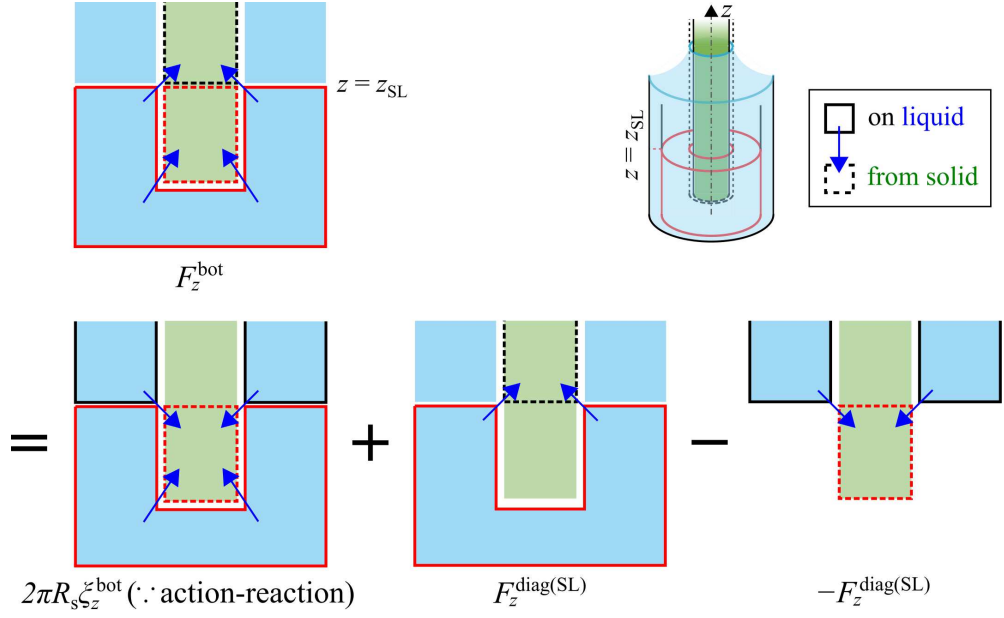}
 \caption{
Schematic of the extraction of the $z$-direction force from the 
solid on the fluid in the red control volume around 
the bottom considering the force distribution in 
Fig.~\ref{Fig:distribtution}.}
\label{Fig:forcebalance}
\end{figure}
\par
We further investigate the curvature dependence of wetting behavior with the calculations of $\xizcl$ and the interfacial tensions.
%curvature dependence of the solid-liquid interfacial tension, 
We start from the extraction of the upward force $F_{z}^\mathrm{bot}$
exerted from the solid on the liquid in the red control volume (CV) around the bottom illustrated in the top-left panel of Fig.~\ref{Fig:forcebalance}.
%as the reaction force from the 
%liquid on the solid, \ie 
Note that we evaluate the upward force $F_{z}$
on the corresponding liquid from the solid as follows: the positive direction for $F_{z}$ is $+z$-direction and is opposite to that for $\xi_{z}$ (force per length) in the $-z$-direction on the corresponding solid from the liquid.
The top face of the CV at $z=z_\mathrm{SL}$ is sufficiently away from both the bottom of the solid and the contact line, %from the solid bottom and contact line.
where the liquid density $\rho$ near the solid is constant in the $z$-direction satisfying
\begin{equation}
  \frac{\ptl \rho}{\ptl z} = 0 \quad 
  \mathrm{for}
  \quad
  \left(r-\Rs\right)^{2} + (z-z_\mathrm{SL})^{2}
  \leq (r_\mathrm{c}^\mathrm{sf})^{2},
  \label{eq:ptlrho_ptlz=0}
\end{equation}
with $r_\mathrm{c}^\mathrm{sf}$ the cut-off distance for solid-fluid interaction.
The force of present interest $F_{z}^\mathrm{bot}$ comes from the neighbouring solid within the cutoff range, \ie from the red-dotted and black-dotted solid parts as indicated by the blue arrows (top-left panel). 
On the other hand, with the condition in Eq.~\eqref{eq:ptlrho_ptlz=0}, the sum of the upward forces on the liquid parts in the red-solid and black-solid lines from the red-dotted solid part (bottom-left panel) is $2\pi \Rs \xizbot$, 
which is the reaction force on the solid around the bottom indicated in Fig.~\ref{Fig:distribtution}.
From the comparison of the arrows regarding the two, $F_{z}^\mathrm{bot}$ is obtained by adding the missing force and subtracting the unnecessary force as in the bottom panel as 
\begin{align}
    \nonumber
    F_{z}^\mathrm{bot} 
    &= 
    2\pi \Rs \xizbot 
    + 
    F_{z}^\mathrm{diag(SL)} - (-F_{z}^\mathrm{diag(SL)})
    \\
    &= 
    2\pi \Rs \xizbot 
    + 
    2F_{z}^\mathrm{diag(SL)}
    \label{eq:Fzbot_org}
\end{align}
where the two ``diagonal" forces denoted by $F_{z}^\mathrm{diag(SL)}$ and 
$-F_{z}^\mathrm{diag(SL)}$ have an opposite sign with the same absolute value due to the symmetry under the  condition in Eq.~\eqref{eq:ptlrho_ptlz=0}. 
\par
The value of unknown $\FzdiagSL$ now must be determined. 
%to derive the SL interfacial tension from the force balance on the liquid in red-solid CV in Fig.~\ref{Fig:forcebalance} including $\xizbot$ in  Eq.~\eqref{eq:Fzbot_org}. 
Although this $\FzdiagSL$ can be obtained directly by MD simulations based on the definition, in the special case where the solid is so smooth compared to the length-scale of solid-fluid inter-particle interaction that the density can be considered constant independent of the position as the present solid with the graphene geometry, $\FzdiagSL$ can be analytically expressed by
\begin{equation}
F_{z}^\mathrm{diag(SL)} = -\pi \Rs \usl.
\label{eq:fzdiag_usl}
\end{equation}
The detailed derivation is described in Appendix~\ref{appsec:meanfield}.
Similarly, the diagonal force $F_{z}^\mathrm{diag(SV)}$ on the vapor below a plane $z=z_\mathrm{SV}$ from the solid above the plane can also be analytically formulated by
\begin{equation}
F_{z}^\mathrm{diag(SV)} = -\pi \Rs \usv.
\label{eq:fzdiag_usv}
\end{equation}
%
%\par
%\rem{We validate Eqs.~\eqref{eq:fzdiag_usl} and \eqref{eq:fzdiag_usv} through the comparison of $\xizcl$ with its analytical expression.}
By considering a force balance similar to that illustrated in Fig.~\ref{Fig:forcebalance} and by also assuming that the solid is smooth and the fluid particles are not pinned around the contact line, the downward force on the solid $\xizcl$ per length is analytically given by (see Appendix~\ref{appsec:xizcl} for details)
\begin{equation}
    \xizcl =  - \usl + \usv = (-\usl) - (-\usv),
\label{eq:xizcl_eq_potdif}
\end{equation}
where the final equality %is appended 
shows that $\xizcl$ is the difference in the magnitude between the two interfacial potential energy densities,
considering that %the potential energy densities 
$\usl$ and $\usv$ are both negative as exemplified in Fig.~\ref{Fig:distribtution}.
\par
\begin{figure}
\centering
\includegraphics[width=1.0\linewidth]{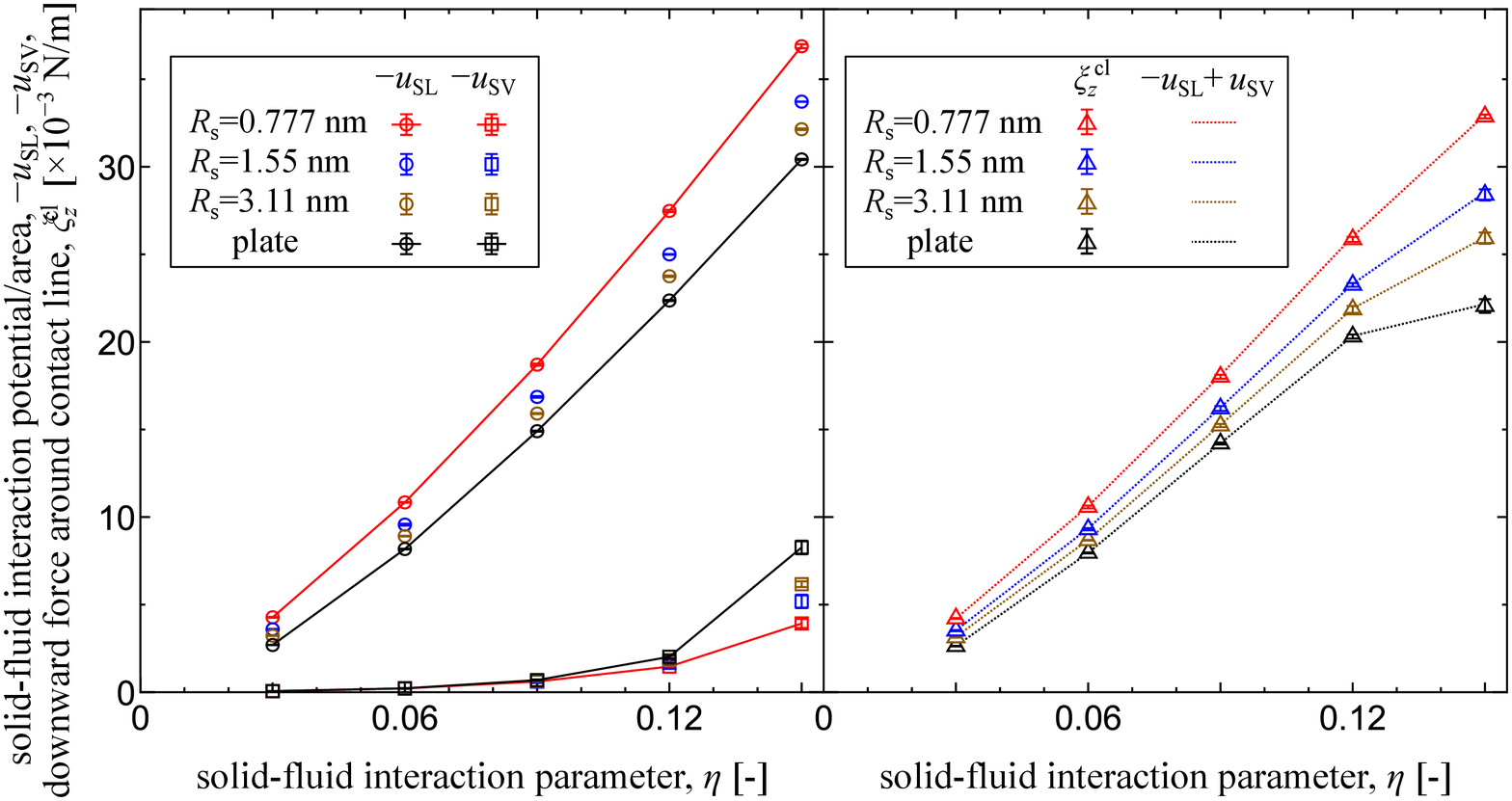}
\caption{
(Left) dependence SL and SV potential 
energy densities $-\usl$ and $-\usv$ 
on the solid-fluid interaction parameter 
$\eta$ for solid plate and solid cylinders
with different radii $\Rs$. (Right) 
Comparison between the downward force $\xizcl$ 
on the solid around the CL and the difference 
of the potential energy density $- \usl + \usv$.
%
%The line for the Wilhelmy relation 
%$\xi_{z} = \glv \cos\theta$ in Eq.~\eqref{eq:Wilhelmy} is also shown.
\label{Fig:usl-usv-xizcl}
}
\end{figure}
Figure~\ref{Fig:usl-usv-xizcl} shows 
the dependence of SL and SV potential energy 
densities $-\usl$ and $-\usv$ 
on the solid-fluid interaction parameter 
$\eta$ for solid plate and solid cylinders
with different radii $\Rs$, and the comparison between 
the downward force $\xizcl$ on the solid around the CL and the difference of potential energy density $- \usl + \usv$.
As easily expected, $-\usl$ and $-\usv$ increased with the increase of $\eta$ as shown in the left panel; however, $-\usl$ and $-\usv$ showed opposite dependence on $\Rs$: $-\usl$ was larger for smaller cylinder radius $\Rs$ whereas $-\usv$ was smaller.
We will discuss this interesting difference later.
In the right panel of Fig.~\ref{Fig:usl-usv-xizcl}, a very good agreement between $\xizcl$ and $- \usl + \usv$
are observed for the whole range of $\eta$ with different radii $\Rs$ in the right panel. This indicates that the force from the solid on the liquid in the CV around the bottom shown as a red concave in Fig.~\ref{Fig:forcebalance} can be properly evaluated  
by Eq.~\eqref{eq:fzdiag_usl} because the present system with solid particles located at the position of graphene was supposed to be sufficiently smooth to meet the condition assumed in the analytical derivation in Appendices~\ref{appsec:meanfield} and \ref{appsec:xizcl}. 
In addition, due to the opposite radius dependence of $-\usl$ and $-\usv$ shown in the left panel, the difference of $\xizcl$ for the smallest radius and that for the plate was as large as about $10\times 10^{-3}$~N/m,
which is comparable to $\glv$. 
In addition, $\xizcl$ was much larger than the difference of $-(\gsl-\gs0)$  and $-(\gsv-\gs0)$ shown later.
%observed in Fig.~\ref{fig:gsl_gsv}.
%
%Considering the action-reaction 
%relation between the solid and fluid, the upward force on that 
%liquid parts is equal to $2\pi \Rs \xizbot$ (top-middle panel).
%To extract the total $z$-direction force from the dotted-red liquid part on the solid-red and solid-black solid parts, one has to add the force from the red-dotted liquid part on the solid-black solid part (bottom-middle panel) and 
%has to subtract that from the dotted-black liquid part on the red-solid solid part (bottom-right panel). 
%From the symmetry, the two forces have the same absolute value with an opposite sign, and we define the force 
%by $F_{z}^\mathrm{diag(SL)}$ and $-F_{z}^\mathrm{diag(SL)}$, respectively.
%
%and 
%
\par
%, indicating that the resulting relative interfacial tensions $\gsl - \glv$ and $\gsv-\gs0$ using $\usl$ and $\usv$ in Eqs.~\eqref{eq:gsl_final} and \eqref{eq:gsv_final}, respectively, are feasible.
%As $\Fzbot$ in Eq.~\eqref{eq:Fzbot_org} including $\FzdiagSL$ was properly evaluated,  
We now examine the relative SL interfacial tension by using the total static force balance including the force  $F_{z}^\mathrm{bot}$ on the red-solid CV in Fig.~\ref{Fig:forcebalance}: 
\begin{equation}
2\pi \int_{\rsf}^{\rcv} \drom r\: \tau_{zz}(r,\zsl) r
+
\pi \rcv^{2} \plblk
+ 
F_{z}^\mathrm{bot}
=0,
\label{eq:balance_bot}
\end{equation}
where the 1st and 2nd terms of the LHS are the fluid stress integral 
on the top face and pressure from the bottom, respectively. Note that side boundary is set at the bulk satisfying $\tau_{rz}=0$, 
and also that the fluid stress is zero on the faces around the concave %in contact with the solid 
under the present stress definition.~\cite{Rowlinson1982,Yamaguchi2019,Imaizumi2020,Kusudo2021} 
The first term of the LHS is related to the relative SL interfacial tension $\gsl - \gs0$ with extended Bakker's equation applied for a cylindrical SL interface by (see Appendix~\ref{appsec:extbakker_cylinder})
\begin{align}
%\nonumber
\gamma_{\text{SL}}-\gamma_{\text{S0}}
&=
%\frac{1}{\rsf}\int_{\rsf}^{\rcv}\drom r %\left[\tau_{zz}(r,\zsl)-\tau_\mathrm{L}^\mathrm{blk}\right]r 
%\\ &=
\frac{1}{\rsf} \left[
\int_{\rsf}^{\rcv} \drom r\:
\tau_{zz}(r,\zsl) r
+
\frac{ (\rcv^{2}-\rsf^{2})\plblk}{2}
\right].
\label{eq:Bakker-SL_c}
\end{align}
%
%where $\plblk = -\tau_\mathrm{L}^\mathrm{blk}(=\mathrm{const.})$ is used %for the second equality. 
Thus, from  Eqs.~\eqref{eq:balance_bot}, \eqref{eq:Fzbot_org} and \eqref{eq:Bakker-SL_c}, $\gsl - \gs0$ results in 
\begin{equation}
    \gsl - \gs0
    =
    - 
    \frac{\rs}{\rsf} \xizbot 
    - \frac{\rsf}{2} \plblk 
%    + \frac{\rs}{\rsf} \usl,
    - \frac{F_{z}^\mathrm{diag(SL)}}{\pi\rsf},
    \label{eq:gsl_gen}
\end{equation}
where the pressure $\plblk$ was measured in MD systems as the normal force per area exerted on the bottom potential wall for $r\leq r_\mathrm{CV}$. 
%
%
%Note that $\usl$ is negative as shown in Fig.~\ref{Fig:distribtution}.
%\add{Note that In case the distribution is sparse, ... }
%
Under the condition that Eq.~\eqref{eq:fzdiag_usl} holds for $\FzdiagSL$ as in the present systems %indicated in Fig.~\ref{Fig:usl-usv-xizcl},  
Eq.~\eqref{eq:Fzbot_org} is further rewritten by
\begin{equation}
    F_{z}^\mathrm{bot} = 2\pi \Rs \left(\xizbot - \usl\right),
    \label{eq:Fzbot_fin}
\end{equation}
and Eq.~\eqref{eq:gsl_gen} writes
\begin{equation}
    \gsl - \gs0
    =
    - 
    \frac{\rs}{\rsf} \xizbot 
    - \frac{\rsf}{2} \plblk 
    + \frac{\rs}{\rsf} \usl.
%    - \frac{F_{z}^\mathrm{diag(SL)}}{\pi\rsf}
    \label{eq:gsl_final}
\end{equation}
%
%
%\par
Similarly, the relative solid-vapor interfacial tension $\gsv-\gs0$ writes
\begin{equation}
    \gsv - \gs0
=
    \frac{\rs}{\rsf} 
    \xiztop
    - \frac{\rsf}{2} \pvblk 
    + \frac{\rs}{\rsf} \usv.
    \label{eq:gsv_final}
\end{equation}
Note that Eqs.~\eqref{eq:gsl_final} and \eqref{eq:gsv_final} are equivalent to the derivation for the quasi-2D Wilhelmy plate~\cite{Imaizumi2020} except the point that the radii of the solid surface area $\Rs$ and solid-fluid interface area $\rsf$ are different for the present Wilhelmy-cylinders. Also note that the meniscus shape including the contact angle does not explicitly appear in Eqs.~\eqref{eq:gsl_final} and \eqref{eq:gsv_final}.
\par
\begin{figure}
\includegraphics[width=0.8\linewidth]{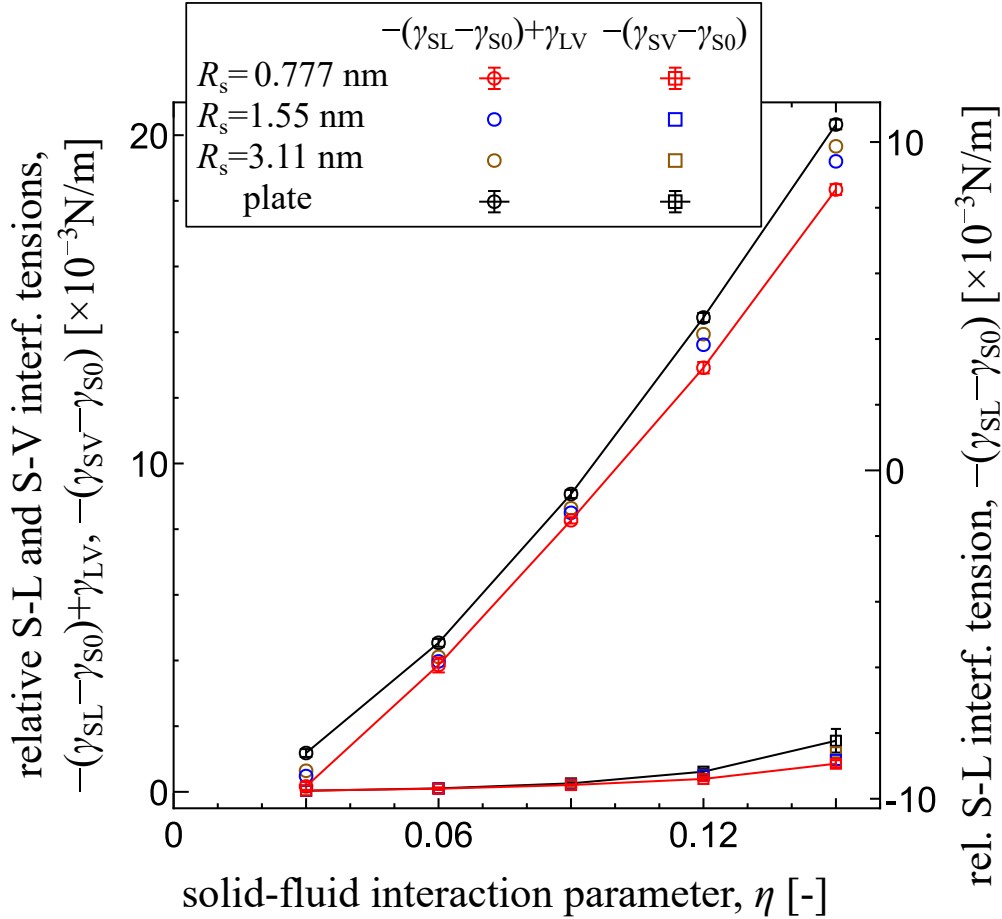}
\caption{Dependence of the SL and SV interfacial tensions on the radius of solid cylinders with different wettability.
}
\label{fig:gsl_gsv}
\end{figure}
Figure~\ref{fig:gsl_gsv} shows the relative SL and SV interfacial tensions calculated by Eq.~\eqref{eq:gsl_final} and \eqref{eq:gsv_final} for different solid radii $\Rs$ with various wettability parameter $\eta$. For a direct comparison between the two interfacial tensions, the values $-(\gsl-\gs0)+\glv$ and $-(\gsv-\gs0)$ are shown with the left vertical axis, which correspond to the works of adhesion:
\begin{equation}
    \Wsl \equiv - (\gsl - \gs0) + \glv 
%    = 
%    -\usl - (-T\Delta s_\mathrm{SL}),
\label{eq:Wsl_gsl_glv}
\end{equation}
and
\begin{equation}
    \Wsv \equiv - (\gsv - \gs0),
%    =
%    -\usv - (-T\Delta s_\mathrm{SV})
\label{eq:Wsv_gsv}
\end{equation}
where the value of $\glv = 9.79\times 10^{-3}$~N/m obtained in our previous study~\cite{Imaizumi2020} was used.
For a flat interface, they are defined as the minimum works 
needed to strip the liquid and vapor off the flat solid surface, respectively 
under constant temperature and pressure condition.~\cite{Yamaguchi2019,Bistafa2021}
Note that both works of adhesion $\Wsl$ and $\Wsv$ are positive, and 
%,
%(except  purely repulsive virtual solid wall), 
%and from Young's equation~\eqref{eq:Young}, their difference is related to the contact angle $\theta_\mathrm{YD}$ estimated by the Young-Dupr\'{e} equation as 
%
%\begin{equation}
%\cos \theta_\mathrm{YD} = \frac{\Wsl-\Wsv}{\glv} - 1.
%\label{eq:Young-Dupre}
%\end{equation}
%
we will discuss about them later from a viewpoint of the free energy as well as the curvature effects. 
With the decrease of the radius $\Rs$, both $-(\gsl-\gs0)$ and $-(\gsv-\gs0)$ became smaller, and the dependence was more remarkable for larger 
%the solid-fluid interaction parameter 
$\eta$ value.
For the smallest 
cylinder with $\Rs=0.777$~nm, $-(\gsl-\gs0)$  was about $2\times 10^{-3}$~N/m smaller than that of the flat plate. However, 
$-(\gsv-\gs0)$ was also reduced with the decrease of $\Rs$, and this resulted in the rather small dependence of the contact angle on the radius shown in Fig.~\ref{Fig:contact_angle}.
\par
\begin{figure}
\centering
\includegraphics[width=0.7\linewidth]{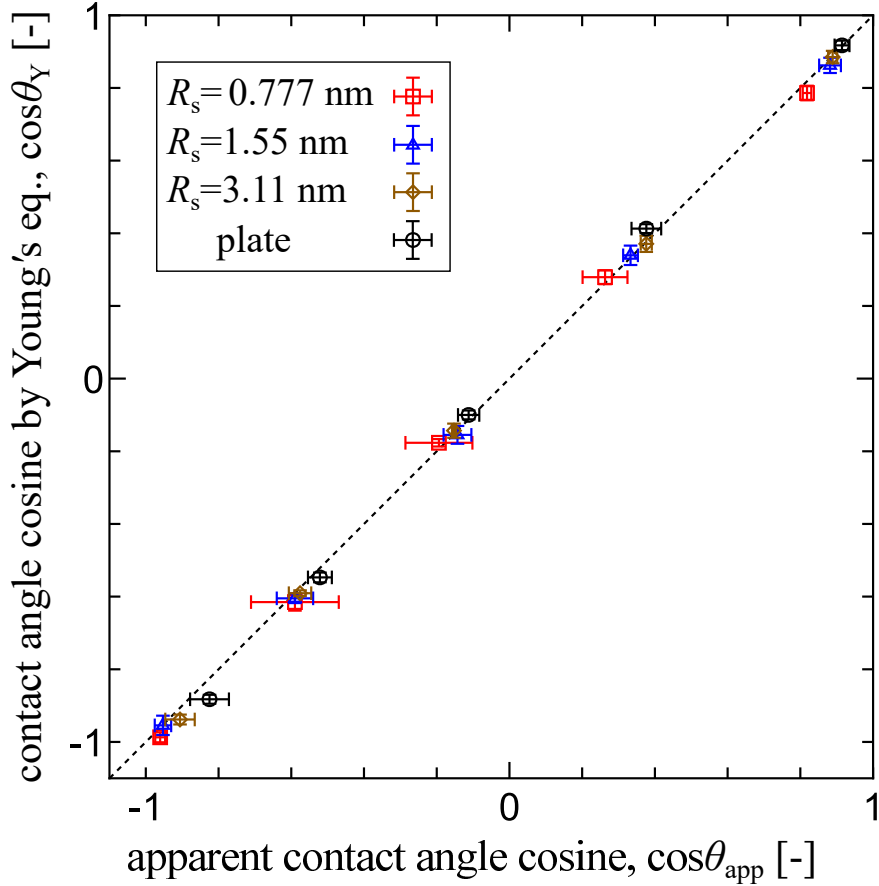}
\caption{
Comparison between the apparent contact angle $\cos \theta_\mathrm{app}$ and that estimated by Young's equation $\cos \theta_\mathrm{Y}$ using the interfacial tensions obtained by the mechanical route. The value of $\eta$ ranges from $0.03$ to $0.15$ for each cylinder radius $\Rs$ and plate.
}
\label{Fig:Young-Dupre}
\end{figure}

\subsection{Applicability of Young's equation}
\label{subsec:Young}
Using the relative interfacial tensions $\gsl - \gs0$ and $\gsv-\gs0$ obtained in the previous subsection, we examined whether Young's equation holds for the present system with a curved solid surface. Figure~7 shows the comparison between the apparent contact angle cosine  $\cos \theta_\mathrm{app}$ in Fig.~\ref{Fig:contact_angle} determined from the meniscus shape and that estimated by Young's equation~\eqref{eq:Young} defined by
\begin{align}
\nonumber
\cos \theta_\mathrm{Y} &=
\frac{\gsv - \gsl}{\glv} \equiv
\frac{(\gsv-\gs0) - (\gsl-\gs0)}{\glv}
\\
& =
\frac{\Wsl-\Wsv}{\glv} - 1
\label{eq:Young_cos}
\end{align}
using the interfacial tensions obtained above via the mechanical route. Note that $\glv$ was set constant considering that its curvature dependence appeared only for a radius of curvature smaller than about $3\sigma_\mathrm{ff}$  for the LJ fluid,~\cite{Yaguchi2010} which is smaller than that in the present study. In addition, it has been shown that $\glv$ consistent with Young's equation should be defined at a position excluding the adsorption layers around the SL interface,~\cite{Yamaguchi2019} at which the radius of curvature of LV interface is sufficiently large and the curvature effect is negligible.
For the whole range of $\eta$ values and radii $\Rs$ tested, $\cos \theta_\mathrm{app}$ and $\cos \theta_\mathrm{Y}$ agreed very well, and this indicates that Young's equation holds for the present systems with curved solid surfaces without pinning if the solid-related relative interfacial tensions $\gsl-\gs0$ and $\gsv-\gs0$ are properly evaluated via a mechanical route. %
\subsection{Discussion}
\begin{figure}
\centering
\includegraphics[width=0.6\linewidth]{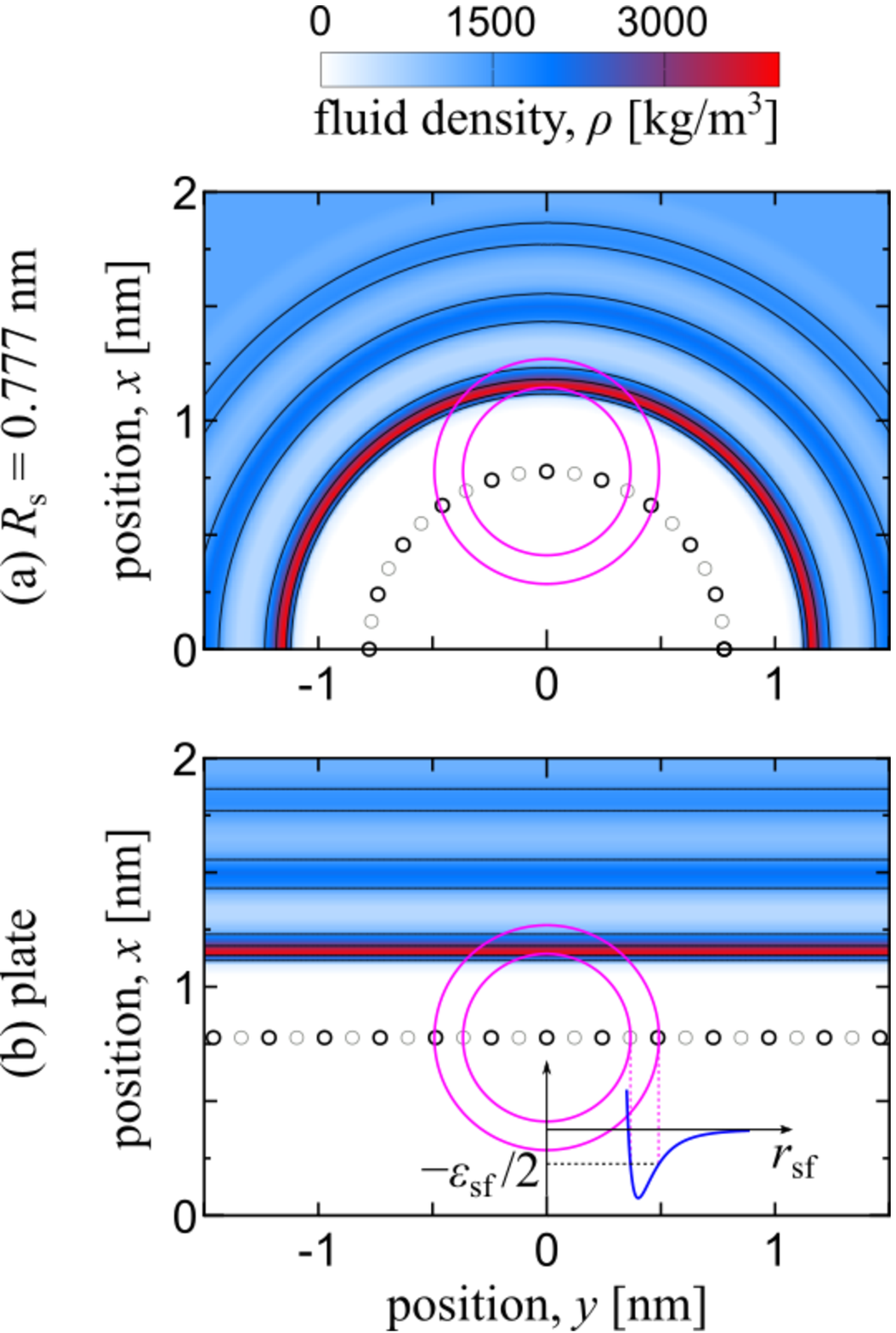}
\caption{
Comparison of the fluid-density around a solid-liquid interface between the (a) cylinder with $\Rs=0.777$~nm and (b) flat plate, with $\eta=0.15$. Positions of the solid particles are shown with small black and gray circles, and the distance range of $r_\mathrm{sf}$ from a solid particle satisfying interaction potential $\Phi_\mathrm{sf}(r_\mathrm{sf})\leq -\epsilon_\mathrm{sf}/2$ is depicted by magenta circles.
}
\label{Fig:sldens-comp}
\end{figure}
A question arises: why was the curvature dependence of the relative interfacial tensions $\gsl-\gs0$ and $\gsv-\gs0$ shown in Fig.~\ref{fig:gsl_gsv} not so large %, and the resulting contact angle consine $\cos \theta$, 
% and \ref{Fig:contact_angle}, 
compared to the contact line force $\xizcl=-\usl + \usv$ seen in Fig.~\ref{Fig:usl-usv-xizcl}? More specifically, why did $-\usl$ in the left panel of Fig.~\ref{Fig:usl-usv-xizcl} and $-(\gsl-\gs0)$ in Fig.~\ref{fig:gsl_gsv} show opposite dependence on $\Rs$?
%
%We start from the 
%\par
To examine the curvature dependence of $-\usl$, we evaluated the density field around the solid-liquid interface, where we have carried out an additional simulation with a solid cylinder with a chiral index (60,0) ($\Rs = 2.33$~nm).
Figure~\ref{Fig:sldens-comp} shows the comparison of the fluid-density around 
a solid-liquid interface between the (a) cylinder with $\Rs=0.777$~nm and (b) flat plate, with $\eta=0.15$. Positions of the solid particles are shown with small black and gray circles, where the particles with the same color are at the same height $z$,  and the distance range of $r_\mathrm{sf}$ from a black solid particle satisfying interaction potential $\Phi_\mathrm{sf}(r_\mathrm{rf})\leq -\epsilon_\mathrm{sf}/2$, as indicated by the blue potential graph, is depicted by magenta circles. 
As seen in this figure, high density region in red corresponding to the fluid first adsorption layer are included in this distance range more for $\Rs = 0.777$~nm, and this results in the higher $-\usl$ for smaller $\Rs$ indicated in the left panel of Fig.~\ref{Fig:usl-usv-xizcl} because the fluid particles in this distance range have the main contribution. Similarly, one can expect smaller $-\usv$ for smaller $\Rs$ because the fluid particles at the SV interface are simply adsorbed on the solid surface as a single particle, \ie these fluid particles do not form adsorption layers as those around the SL interface observed in Fig.~\ref{Fig:distribtution}, and such single particle is subject to weaker interaction potential from the curved solid than from the flat solid.
\par
\begin{figure}
\centering
\includegraphics[width=0.8\linewidth]{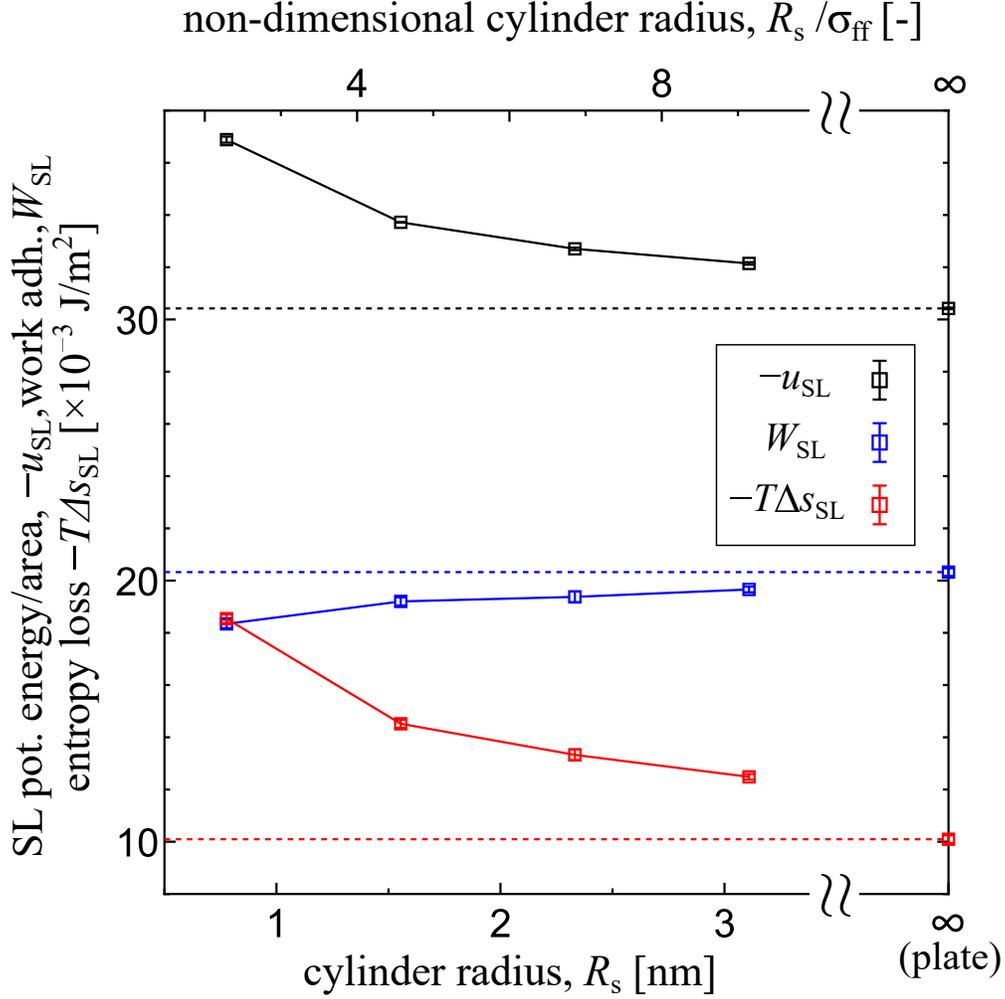}
\caption{
Solid-liquid interfacial potential energy per area $-\usl$, the work of adhesion $\Wsl$, and the entropy loss $-T\Delta s_\mathrm{SL}$ calculated by Eq.~\eqref{eq:Wsl_usl_Tdsl}
for various cylinder radius $\Rs$ with a solid-fluid interaction parameter $\eta = 0.15$.}
\label{Fig:rad-usl-wsl-tdssl}
\end{figure}
Different from the average SL interaction potential $-\usl$, the relative interfacial tension is considered the interfacial free energy per area, \ie  $\gsl-\gs0$ is related to the SL work of adhesion $\Wsl$ in Eq.~\eqref{eq:Wsl_gsl_glv}, and it consists of internal energy and entropy terms:~\cite{Surblys2018,Bistafa2021}
\begin{equation}
    \Wsl 
    =
    - (\gsl - \gs0) + \glv 
    = 
    -\usl - (-T\Delta s_\mathrm{SL}),
    \label{eq:Wsl_usl_Tdsl}
\end{equation}
% %
% and
% %
% \begin{equation}
%     \Wsv 
%     =
%     -\usv - (-T\Delta s_\mathrm{SV}),
% \end{equation}
%
where $-T\Delta s_\mathrm{SL}$ 
%and $\Delta s_\mathrm{SV}$ are 
is due to the entropy ``loss" $-\Delta s_\mathrm{SL}$ induced by the density increase in the adsorption layers of the SL interface.
%, respectively.
Figure~\ref{Fig:rad-usl-wsl-tdssl} shows the solid-liquid interfacial potential energy per area $-\usl$, the work of adhesion $\Wsl$ and the entropy loss $-T\Delta s_\mathrm{SL}$ in Eq.~\eqref{eq:Wsl_usl_Tdsl} for various cylinder radius $\Rs$
with $\eta = 0.15$. Note that the additional data for a cylinder radius $\Rs = 2.33$~nm were also shown, and the radius for the plate was set as $\Rs=\infty$.
%
%As Equation~\eqref{eq:Wsl_usl_Tdsl} indicates that
A larger entropy loss exceeding the potential gain in $-\usl$ %was observed due to the  solid curvature, and this 
resulted in the opposite curvature dependence between the SL interfacial potential energy $-\usl$ and the relative interfacial tension $-(\gsl-\gs0)$ seen in Figs.~\ref{Fig:usl-usv-xizcl} and \ref{fig:gsl_gsv}. In addition, it is indicated that the curvature effect could remain even for a relatively large cylinder radius $\Rs$ larger than about $10\sigma_\mathrm{ff}$, which is much larger than the radius dependence range of $\glv$ reported for the LJ droplets.~\cite{Yaguchi2010}
%Indeed, interfacial entropy can be evaluated.~\cite{Surbly2018,Bistafa2021}
%
%\par
% Regarding the evaluation of the forces from the solid on a liquid part in a specific region, we used the analytical expression with $\usl$ or $\usv$ considering the condition that the present solid surface were sufficiently smooth, but a direct extraction of such force from MD would be possible. A direct extraction of the pinning force would be also possible~\cite{Kusudo2019} to  further examine the applicability of Young's equation. On the other hand, for the calculation of the relative SL interfacial tension $\gsl - \gs0$ in Eq.~\eqref{eq:gsl_gen}, only the force on the solid bottom $\xizbot$, diagonal force $\FzdiagSL$ and bulk pressure $\plblk$ are needed, and a large system with meniscus as in the present study is not necessarily needed, meaning that $\gsl - \gs0$ could be calculated in a smaller system with a low computational cost and this would further improve the applicability of the present nano-Wilhelmy method.
%enables the calculation of  
% computationally demanding calculations of the local stress distributions and the thermodynamic integration.
%
%
\section{CONCLUDING REMARKS}
In this study, we successfully extracted the SL and SV interfacial tensions of a simple Lennard-Jones fluid around a solid cylinder with a nanometer-scale diameter by extending the theoretical nano-Wilhelmy equations for a quasi-two-dimensional flat solid plate from our previous study. 
The SL and SV interfacial tensions were calculated from the integral of the normal-stress in the wall-tangential direction by considering the mechanical force balances on control volumes set around the bottom and top of the solid cylinder  subject to the fluid stress and external force from the solid, where the local force on the solid around the contact line expressed by these external forces agreed well with the analytical expression.
The theoretical contact angle calculated by Young's equation using these interfacial tensions agreed well with the apparent contact angle estimated by the analytical solution fitted to the meniscus shape, showing that Young's equation holds even for menisci around solids with nanoscale curvature if the interfacial tensions are properly evaluated.
It was also shown that the curvature dependence of the SL and SV interfacial tensions as the free energy was different from that of the corresponding interaction potential energies as a part of the internal energy, which explains the weak curvature dependence of the contact angle in the present results.
\par
The accurate calculation of the interfacial tensions on curved surfaces could explain the unique wetting behavior of water on carbon nanotubes.~\cite{Homma2013, Imadate2018} In addition, it should enable the exploration of the Tolman equation for the solid-related interfaces. Related to this, an interesting future target is the interfacial tensions inside the curved interface from a mechanical route,  which should enable the analysis of nano-confinement effects or nanoscale capillary as well, \eg in carbon nanotubes.
%however; another framework would be needed to extract the stress integral inside the confined space for this extension.

%In this work, we proposed theoretical model for the local forces exerted on a quasi-2D smooth solid plate immersed into a liquid pool of a simple liquid based on a mean-field approach and on the connection between $\gsl$ or $\gsv$ and stress distribution at　each　interface. 
%The analytical forms derived for the local force around the CL and total force on the solid plate agreed very well with the MD simulations, which indicated the validity of the  rigorous formulation based on the equilibrium force balance.
%
% In the present study, we set up an axi-symmetric system with a cylinder dipped into a liquid pool with additional cylindrical side wall because we tried to simultaneously calculate the interfacial tensions and the contact angle from the axi-symmetric meniscus shape for the connection with Young's equation. However, 
% for the calculation of the solid-liquid interfacial tension, for instance, through the extraction of the stress integral of $\tau_{zz}$ in Eqs.~\eqref{eq:Bakker-SL_c}-\eqref{eq:gsl_final}, would be possible even in a system with a cylinder vertically dipped into a liquid pool with periodic boundary conditions in the horizontal directions, \ie without cylindrical side wall: if the static force balance similar to that shown in Fig.~\ref{Fig:forcebalance} were properly considered.
%
%\add{Todo}
%
\begin{acknowledgments}
%We thank Konan Imadate  for fruitful discussion. 
T.O., H.K. 
and Y.Y. were supported by JSPS KAKENHI grant (Nos. JP18K03929, 
JP20J20251
and JP18K03978), Japan, respectively. 
Y.Y. was also supported by JST CREST grant (No. JPMJCR18I1), Japan.
\end{acknowledgments}
%
%\newline
\vspace{5mm} \par \noindent
\textbf{DATA AVAILABILITY}
%\newline
\par
The data that support the findings of this study are available from the corresponding author
upon reasonable request.
\vspace{5mm} \par \noindent
\textbf{AUTHOR DECLARATIONS}
\newline
\textbf{Conflict of Interest}
\par
The authors have no conflicts to disclose.
\appendix
\section{Extended Bakker's equation for cylinder
\label{appsec:extbakker_cylinder}
}
\begin{figure}
\centering
\includegraphics[width=0.7\linewidth]{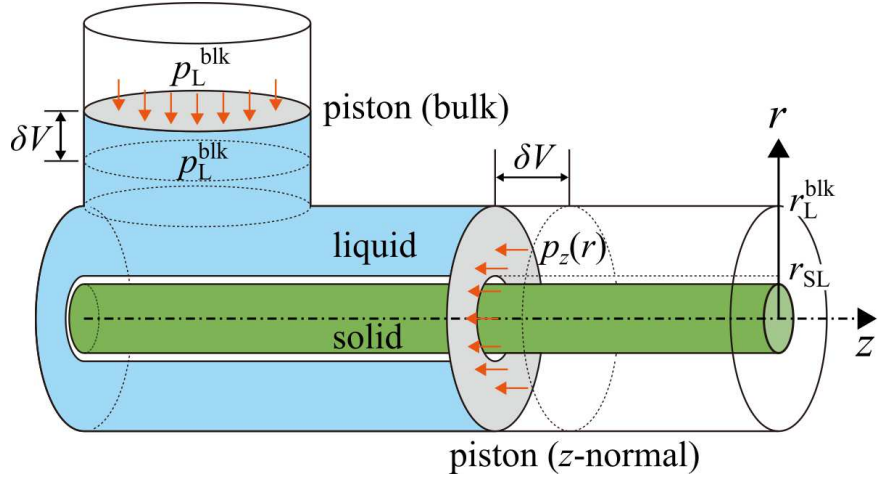}
\caption{
Thought experiment for the connection between relative solid-liquid interfacial tension and pressure distribution around the solid cylinder.
\label{fig:bakker_cylinder}
}
\end{figure}
We formulate the relative solid-liquid (SL) interfacial tension
relative to the solid-vacuum (S0) $\gsl-\gs0$ 
between cylindrical solid surface and liquid through the thought experiment shown in Fig.~\ref{fig:bakker_cylinder}. The side piston normal to the cylinder axis $z$ is in contact only 
with the liquid, \ie the inner radius is at $r_\mathrm{SL}$ set at the limit radius nearest 
to the solid that the fluid particle can reach, whereas the outer radius $\rlblk$ is at liquid bulk sufficiently away from the SL interface. 
The axis-normal stress defined by
\begin{equation}
    \tau_{zz}(r) \equiv  - p_z (r)
    \label{eq_app:tauzz=-pz}
\end{equation}
is a function of radial position $r$, and it satisfies
\begin{equation}
p_{z}(r)= \plblk=\text{const.}\quad (r \geq \rlblk).
\label{eq:tauzz_at_r>rblk}
\end{equation}
On the other hand, the top piston is set at the liquid bulk,
on which homogeneous pressure %$p_\mathrm{N}$ 
identical to the bulk pressure $\plblk$ 
is exerted. 
%\ie the radial normal stress 
%$\tau_{rr}\equiv p_\mathrm{N}$ 
%on the piston is
%
%
%\begin{equation}
%\tau_{rr}=- p_\mathrm{N}= -\plblk =\text{const.}
%\label{eq:taurr}
%\end{equation}
%
We suppose a virtual infinitesimal displacement $\delta z$ of the side piston with a simultaneous downward displacement of the top piston so that the 
liquid volume may not change.
If the operation is quasi-static at a constant temperature, the change of the Helmholtz free energy $\delta F$ is equal to the work $\delta W$ exerted on the system given by
\begin{align}
\label{eq:deltaF-mechanical_c}
%\begin{split}
\delta F
=
\delta W
&= 
\plblk \delta V - 2\pi \delta z \int_{r_\mathrm{SL}}^{ \rlblk} \drom r\: p_z (r)r 
%\\
%&= 2\pi \delta  \int_{r_\text{B}}^{ r_\text{T}} \left[p_{\text{N}}-p_z (r)\right]r \text{d}r
%\end{split}
\end{align}
where the volume increase and decrease $\delta V$
due to the motions of side and top pistons, respectively writes
\begin{equation}
\delta V=2\pi \delta z \int_{r_\mathrm{SL}}^{ \rlblk} \drom r\: r
=
 \pi\left[ (\rlblk)^{2} - r_\mathrm{SL}^{2}\right] \delta z
\end{equation}
The free energy change $\delta F$ in Eq.~\eqref{eq:deltaF-mechanical_c} 
is uniquely defined as long as $\rlblk$
is set in the bulk satisfying Eq.~\eqref{eq:tauzz_at_r>rblk}.
From a macroscopic point of view, 
the SL interface is increased and S0 interface is reduced 
with this operation. By assuming that the SL interface is 
at $r_\mathrm{SL}$, it follows that
\begin{equation}
\label{eq:deltaF-thermodynamic-SL_c}
\delta F=2\pi r_\mathrm{SL} (\gsl-\gs0) \delta z 
\end{equation}
By equating Eqs.~\eqref{eq:deltaF-mechanical_c} and 
\eqref{eq:deltaF-thermodynamic-SL_c}, the following relation 
is derived as extended Bakker's equation for a cylindrical 
SL interface:
\begin{align}
\gamma_{\text{SL}}-\gamma_{\text{S0}}
% &=
% \frac{1}{r_\mathrm{SL}}\int_{r_\mathrm{SL}}^{\rlblk} \drom r\:
% \left[\plblk - p_{z}(r)\right]r
%\nonumber
% \\
=
\frac{1}{r_\mathrm{SL}}\left[
\int_{r_\mathrm{SL}}^{\rlblk} \drom r\:
\tau_{zz}(r) r 
+  \frac{\plblk\left[(\rlblk)^{2} - r_\mathrm{SL}^{2}\right]}{2}
\right],
\label{eq:Bakker-SL_c_app}
\end{align}
where Eq.~\eqref{eq_app:tauzz=-pz} is used as well.
Equation~\eqref{eq:Bakker-SL_c_app}
means that the relative SL interfacial tension 
is obtained with the stress integral, bulk pressure and the 
interface position. 
\par 
The relative solid-vapor (SV) interfacial tension is expressed as well by
\begin{align}
\label{eq:Bakker-SV_c_app}
\gsv-\gs0
&=
\frac{1}{r_\mathrm{SV}}\left[
\int_{r_\mathrm{SV}}^{\rvblk} \drom r\:
\tau_{zz}(r) r 
+  \frac{\pvblk\left[(\rvblk)^{2} - r_\mathrm{SV}^{2}\right]}{2}
\right].
\end{align}
\section{Extraction of the interaction force between the liquid and solid across a $z$-normal plane at a SL interface.
\label{appsec:meanfield}
}
\begin{figure}
\centering
\includegraphics[width=0.5\linewidth]{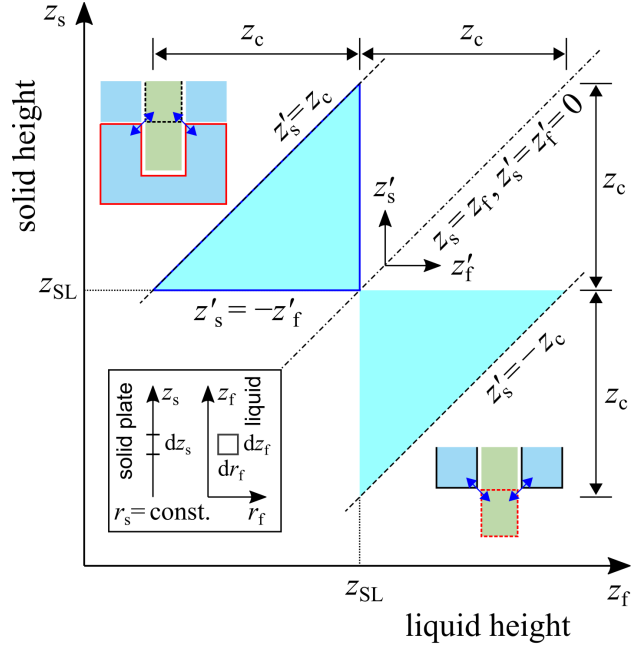}
\caption{
Region for the double integral of the mean field to calculate the interaction between the solid cylinder above $z_\mathrm{s}=z^\mathrm{SL}$ and liquid below $z_\mathrm{f}=z^\mathrm{SL}$ (top-left triangle in solid blue line). 
The geometrical relation is shown in the inset. The cutoff distance $z_\mathrm{c}$ for $|\zf - \zs|$ depends on the relative radial position $\rf - \rs$.
\label{fig:dbl-integral}
}
\end{figure}
We derive the interaction force %$F_\mathrm{z}^\mathrm{diag}$ (denote by $F_\mathrm{z}^\mathrm{diag(SL)}$ here) 
$F_\mathrm{z}^\mathrm{diag(SL)}$
exemplified in the bottom panels of Fig.~\ref{Fig:forcebalance}, namely, the interaction force between the solid above a $z$-normal plane $z=z_\mathrm{SL}$ and liquid below that plane. The plane $z=z_\mathrm{SL}$ is at a height around which the fluid density is independent of the position $z$ because the plane is sufficiently away from the contact line and the bottom of the cylinder.
\par
Taking into account that the solid is supposed to be smooth for the fluid particles because the interparticle distance parameters $\sigma_\mathrm{ff}$ and $\sigma_\mathrm{sf}$ are sufficiently large compared to $r_\mathrm{ss}$ between solid particles, the solid-liquid interaction force can be analytically modeled by assuming the mean fields of the fluid and solid.
The mean number density per volume 
$\rnf(\rf)\ (=\rho/m_\mathrm{f})$ of the fluid is 
given as a function of the radial 
position $\rf$ of the fluid, whereas a 
constant mean number density 
per area $\rns$ of the solid  at $r=\rs$ is used considering the present system
with a solid cylinder of zero-thickness without volume; however, the
following derivation can easily be extended for a system with a solid 
with a volume and density per volume in the range $r \leq \rs$.
%\rem{as long as the density is independent of $\zs$}

%
\par
We start from the potential energy on a solid particle at  position $(\rs,\vts,\zs)$ due to a fluid particle at $(\rf,\vtf,\zf)$ both in the cylindrical $r\vartheta z$-coordinate, given by Eq.~\eqref{eq:LJ}. 
Assuming that the fluid particles are homogeneously 
distributed in the azimuth direction $\vartheta$ with a number density $\rnf(\rf,\zf)$ per volume, the mean potential field  
from an infinitesimal fluid volume segment of 
$\mathrm{d}\zf \times 2\pi \rf \drf$ 
on a solid particle 
is defined by using 
$\rnf(\zf,\rf)$
and the mean local potential $\phi(\zpf, \rpf)$
as 
$\rnf(\zf,\xf)\dzf \drf
\cdot
\phi(\zpf, \rpf)$, 
where $\phi(\zpf, \rpf)$ is given by
\begin{equation}
\phi(\zpf, \rpf)% \drom \zf \drom \rf 
\equiv 
%\rnf(\rf,\zf) \cdot \left[
\int_{0}^{2\pi} \Phi_\mathrm{LJ}
\left(r_\mathrm{sf} \right) \rf \drom\vtf
%\right]\drom \zf \drom \rf ,
\end{equation}
where we define
\begin{equation}
r_\mathrm{sf}=
\sqrt{
(\rf \cos \vtf - \rs \cos \vts)^{2} + 
(\rf \sin \vtf - \rs \sin \vts)^{2} + 
\zpf^{2}},\quad
\label{app_eq:def_localphi}
\end{equation}
and 
\begin{equation}
\zpf \equiv \zf-\zs \equiv -\zds,\quad
\rpf \equiv \rf-\rs. % \equiv -\rs{'} 
\label{app_eq:def_relpos}
\end{equation}
Then, the local force in the $z$-direction
$f_{z}^\mathrm{s}(\zpf,\rpf)\dzf \drf \dzs$
exerted on the solid cylinder in 
$[\zs, \zs+\dzs]$ with an area number density of the 
solid particles $\rns(\zs)$ 
%a solid particle 
from the present fluid volume-segment 
is given by:
\begin{align}
%\nonumber
f_{z}^\mathrm{s}(\zs,\zf,\rf)
\dzf \drf \dzs
&=
-
%\frac{\rem{\partial}}{\rem{\partial \zs}}%\left[
\rnf(\zf,\rf)
\frac{\partial \phi(\zpf, \rpf)}{\partial \zs}
\mathrm{d}\zf \mathrm{d}\rf \cdot 
2\pi \rs \rns(\zs) \dzs
%\cdot \rns(\zs)\mathrm{d}\zs
% \\
% &=
% -
% \rns(\zs) \rnf(\zf,\rf)
% \frac{
% \partial \phi(\zpf, \rpf)
% }{
% \partial \zs}
% \drom\zf \drom\rf \drom\zs,
\label{app_eq:forcesegment}
\end{align}
where 
\begin{equation}
f_{z}^\mathrm{s}(\zs,\zf,\rf) = 
- 2\pi \rs \rns(\zs)
\rnf(\zf,\rf)
\frac{
\partial \phi(\zpf, \rpf)
}{
\partial \zs}
\label{app_eq:forcedensity}
\end{equation}
denotes the axial force density on the solid given as a function of $\zs$, $\zf$ and $\rf$.
\par
Since $\Phi_\mathrm{LJ}(r_\mathrm{sf})$ is truncated at the cutoff distance $\rc$ in the present case, 
\begin{gather}
\phi \left(\zpf, \rpf\right) 
=0,\quad
\frac{\ptl \phi\left(\zpf, \rpf\right)}{\ptl \zs} 
= 0
\label{app_eq:phi_limit}
\\
\mathrm{for}
\quad
|\zpf| \geq \sqrt{\rc^{2} - \rpf^{2}} \equiv
\zc(\rpf)
\quad \mathrm{or} \quad 
\rpf \geq \rc
\nonumber
\end{gather}
holds,
where $\zc(\rpf)$ as a function of $\rpf$ denotes the cutoff 
with respect to $\zpf$. % and $\zds$
%in the following with respect to 
This cutoff is not critical as long as $\phi\left(\zpf, \rpf\right)$ quickly vanishes with the increase of $r$, but we continue the derivation including the cutoff for simplicity.
With the definition of $r_\mathrm{SF}$ as the limit that the fluid could reach, it follows that 
\begin{equation}
\rnf=0 
\quad \mathrm{for} \quad
\rf < \rsf.
\end{equation}
In addition, considering that $\phi(\zpf, \rpf)$ is 
an even function with respect to $\zpf$, \ie
\begin{equation}
\phi\left(\zpf, \rpf\right)
=
\phi(-\zpf, \rpf),
\label{app_eq:phi_even}
\end{equation}
it follows for the mean local potential $\phi$ that
\begin{equation}
\frac{\partial \phi(\zpf, \rpf) }{\partial \zs} 
=
-\frac{\partial \phi(-\zpf, \rpf) }{\partial \zs}, 
\label{app_eq:phi_der_oddfunc}
\end{equation}
and
\begin{equation}
\frac{\partial \phi(\zpf, \rpf) }{\partial \zs} 
=
-\frac{\partial \phi(\zpf, \rpf) }{\partial \zf},
\label{app_eq:exchange_zs_zf}
\end{equation}
where Eq.~\eqref{app_eq:def_relpos} is applied for the latter.
This corresponds to the action-reaction relation between 
solid and fluid particles under a simple two-body 
interaction, \ie 
\begin{equation}
f_{z}^\mathrm{f}(\zs,\zf,\rf) 
=
-f_{z}^\mathrm{s}(\zs,\zf,\rf)
=
-2\pi \rs \rns(\zs)\rnf(\zf,\rf)
\frac{\partial \phi(\zpf, \rpf)}{\partial \zf}
\label{app_eq:forcedensity_f}
\end{equation}
holds for the tangential force density on the  fluid $f_{z}^\mathrm{f}$.
\par
Based on these properties, we now derive the analytical expression of the force exerted on the solid above a $z$-normal plane $z=z_\mathrm{SL}$ from the liquid below that plane, \ie the force of interest $F_{z}^\mathrm{diag(SL)}$ given by 
\begin{align}
\nonumber
F_{z}^\mathrm{diag(SL)}
&=
\int_{0}^{\rc} \drom \rpf
%\left[
    %\left(\rho_{V}^\mathrm{f(SL)}(\xpf) 
    \int_{-\zc(\rpf)}^{0}\drom \zpf
%left(
\int_{-\zpf}^{\zc(\rpf)} \drom \zds 
f_{z}^\mathrm{f}
%\right) 
\\
&=
-\int_{0}^{\rc} \drom \rpf
\left[
    %\left(\rho_{V}^\mathrm{f(SL)}(\xpf) 
    \int_{-\zc(\rpf)}^{0}\drom \zpf
\left(
\int_{-\zpf}^{\zc(\rpf)} \drom \zds 
f_{z}^\mathrm{s}
\right) 
\right],
\label{app_eq:def_Fzdiag}
\end{align}
where the double integral in the square brackets corresponds to the top-left region in Fig.~\ref{fig:dbl-integral}.
Let the density $\rnf$ for 
$\zsl - \zc < \zf < \zsl +  \zc$
be given as a unique function of $\rf$ by
\begin{equation}
\rnf(\zf,\rf) = 
\rho_{V}^\mathrm{f(SL)}(\rf).
\label{app_eq:rnf_at_SL_SV}
\end{equation}
Then, it follows for the double integral in the square brackets in Eq.~\eqref{app_eq:def_Fzdiag} %\rem{corresponding to the top-left region in  Fig.~\ref{fig:dbl-integral}}
that
\begin{align}
\nonumber
& \int_{-\zc}^{0} \drom \zpf
\left(
\int_{-\zpf}^{\zc} \drom \zds 
f_{z}^\mathrm{s} \right) 
\\ \nonumber
&=
-2\pi \rs
\int_{-\zc}^{0} \drom \zpf 
\rho_{V}^\mathrm{f(SL)}(\rpf)
\left[
\int_{-\zpf}^{\zc} \drom \zds
\rns(\zs)
\frac{\partial \phi(\zpf, \rpf)}{\partial \zs} 
\right] 
\\ \nonumber    
&=
-2\pi \rs \int_{-\zc}^{0}
\rho_{V}^\mathrm{f(SL)}(\rpf) \drom \zpf
\left\{
\left[\rns(\zs)\phi(\zpf, \rpf)
\right]_{\zds=-\zpf}^{\zc}
-\int_{-\zpf}^{\zc} \drom\zds
\frac{d \rns(\zs)}{d \zs}\phi(\zpf, \rpf)
\right\}
\\ \nonumber    
&=
2\pi \rs \int_{-\zc}^{0} \drom \zpf
\rho_{V}^\mathrm{f(SL)}(\rpf)
\left[
\rns(-\zpf)\phi(-\zpf, \rpf)
+\int_{-\zpf}^{\zc} \drom \zds
\frac{d \rns(\zs)}{d \zs}\phi(\zpf, \rpf)
\right],
\\ \nonumber    
&=
2\pi \rs \int_{-\zc}^{0} \drom\zpf
\rho_{V}^\mathrm{f(SL)}(\rpf)\rns(-\zpf)
\phi(\zpf, \rpf)
\nonumber
\\
& +
2\pi \rs \int_{-\zc}^{0} \drom \zpf
\rho_{V}^\mathrm{f(SL)}(\rpf)
\left[
\int_{-\zpf}^{\zc} \drom \zds
\frac{d \rns(\zs)}{d \zs}\phi(\zpf, \rpf)
\right]
\end{align}
where $\phi(\zc, \rpf)=0$ and 
Eq.~\eqref{app_eq:phi_even}
are used for the 4th equality. 
\par
With an additional assumption of 
\begin{equation}
    \rns =\mathrm{const.},
    \label{eq:rns=const_app}
\end{equation} 
the 2nd term of the right-most HS becomes zero, and it follows
\begin{align}
\nonumber
\int_{-\zc}^{0}
\drom \zpf
\left(
\int_{-\zpf}^{\zc}\drom \zds
f_{z}^\mathrm{s} \right)
&=
2\pi \rs \rns \int_{-\zc}^{0} \drom \zpf
\rho_{V}^\mathrm{f(SL)}(\rpf)
\phi(\zpf, \rpf)
\\
&=
%\frac{2\pi \rs \rns}{2}
\pi \rs \rns
\int_{-\zc}^{\zc} \drom \zpf
\rho_{V}^\mathrm{f(SL)}(\rpf)
\phi(\zpf, \rpf)
\label{app_eq:dint_bl_vert}
\end{align}
considering that $\phi$ is an even function
with respect to $\zpf$ for the second equality. 
By inserting Eq.~\eqref{app_eq:dint_bl_vert} into Eq.~\eqref{app_eq:def_Fzdiag}, 
%the force on the solid from the liquid $-F_{z}^\mathrm{diag(SL)}$ at a fixed radius range $\rpf \in [0, r_\mathrm{c}]$ is obtained by further integrating with respect to $\rpf$ as
it follows
\begin{equation}
    F_{z}^\mathrm{diag(SL)}
    =
    %-\frac{2 \pi \rs \rns}{2} 
    -\pi \rs \rns
    \int_{0}^{\rc} \drom \rpf
    %\left(\rho_{V}^\mathrm{f(SL)}(\xpf) 
    \int_{-\zc(\rpf)}^{\zc(\rpf)}\drom \zpf
    \rho_{V}^\mathrm{f(SL)}(\rpf)
\phi(\zpf, \rpf) 
%\equiv \usl.
\label{app_eq:fzdiagSL}
\end{equation}
Indeed, the RHS of Eq.~\eqref{app_eq:fzdiagSL} can be expressed using the following SL potential energy density 
$\usl$ given by 
\begin{equation}
     \usl \equiv
    \rns\int_{0}^{\rc} \drom \rpf
    %\left(\rho_{V}^\mathrm{f(SL)}(\xpf) 
    \int_{-\zc(\rpf)}^{\zc(\rpf)}\drom \zpf
    \rho_{V}^\mathrm{f(SL)}(\rpf)
\phi(\zpf, \rpf) 
\label{app_eq:def_usl_rev}
\end{equation}
which represents the SL potential energy per 
solid area at the SL interface away both from 
the CL and from the bottom of the solid plate.
With $u_\mathrm{SL}$, Eq.~\eqref{app_eq:fzdiagSL} writes
\begin{equation}
   F_{z}^\mathrm{diag(SL)} 
   =
  -\pi \rs \usl
   \label{app_eq:Fzdiag_usl}
\end{equation}
which corresponds to  Eq.~\eqref{eq:fzdiag_usl}
in the main text.
\par
Similar to $F_\mathrm{z}^\mathrm{diag(SL)}$, the interaction force $F_\mathrm{z}^\mathrm{diag(SV)}$ between the solid above a $z$-normal plane $z=z_\mathrm{SV}$ and the vapor below that plane % set between the top of the solid and the CL as well as away from both 
writes
\begin{equation}
   F_{z}^\mathrm{diag(SV)} 
   =
   -\pi \Rs \usv,
\end{equation}
where the SV potential energy density $\usv$ is 
given by
%
%the assumption in Eq.~\eqref{eq:rns=const_app} 
%given by
\begin{equation}
\usv \equiv
\rns\int_{0}^{\rc} 
\mathrm{d}\rpf\,
    \rho_{V}^\mathrm{f(SV)}(\rpf) 
\int_{-\zc(\rpf)}^{\zc(\rpf)}
\mathrm{d}\zpf\,
    \phi(\zpf, \rpf).
\label{app_eq:def_usv}
\end{equation}
%
%which can be derived in a similar manner.
%
%\par
%
\section{Extraction of the force exerted on the solid around the contact line.
\label{appsec:xizcl}
}
\begin{figure}
\centering
\includegraphics[width=1.0\linewidth]{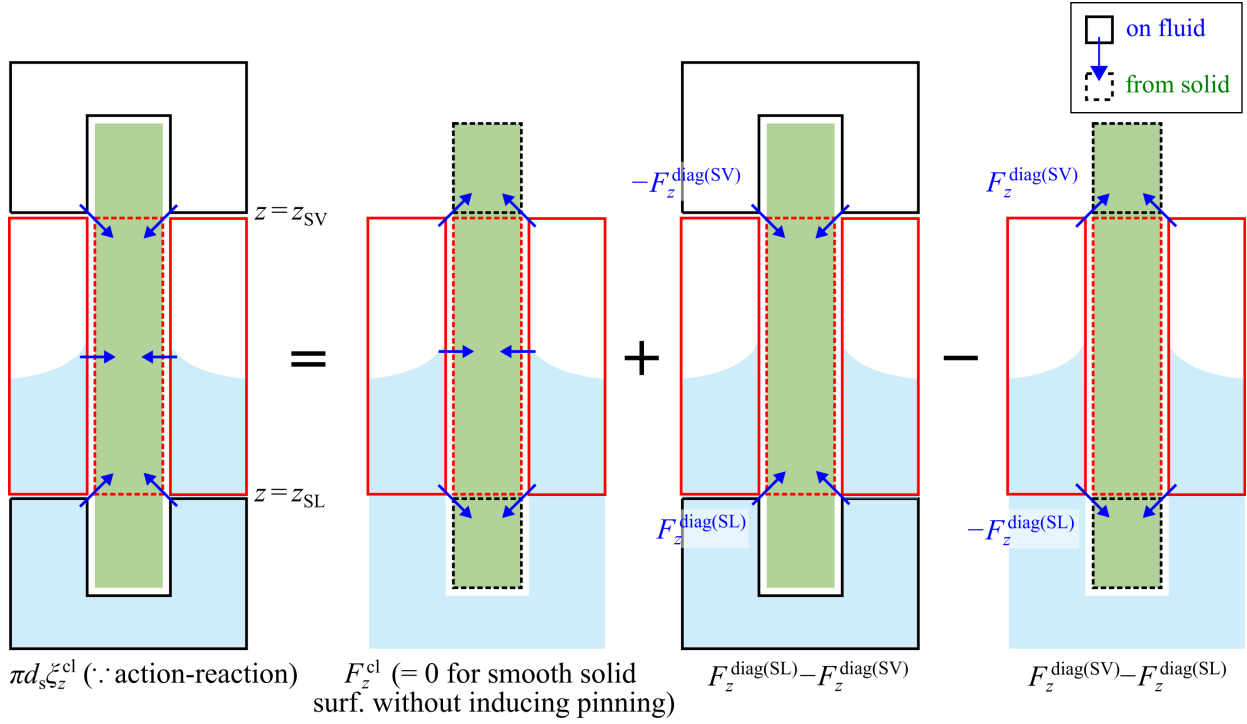}
\caption{
Schematic of the extraction of the $z$-direction force on the solid around the contact line from the fluid.
\label{fig:schematic_CLforce}
}
\end{figure}
We can formulate the downward force on the solid around the contact line $2 \pi \Rs \xizcl$ from the fluid as the reaction force, \ie the upward force on the liquid from the solid, from the force balance similar to that displayed in Fig.~\ref{Fig:forcebalance}.
Let $F_{z}^\mathrm{cl}$ be the force on the liquid around the contact line, $2\pi \Rs \xizcl$ can be obtained by adding 
$\FzdiagSL - \FzdiagSV$ and subtracting $-\FzdiagSL + \FzdiagSV$ as
\begin{equation}
    2\pi \Rs \xizcl = F_{z}^\mathrm{cl} 
    + 2\left(
    \FzdiagSL - \FzdiagSV
    \right)
    \label{app_eq:forcebalance_CL}
\end{equation}
as illustrated in Fig.~\ref{fig:schematic_CLforce}.
%
%Thus, Indeed $F_{z}^\mathrm{cl}$ can easily be obtained directly from MD simulations, but 
In case the solid surface is smooth and flat and no pinning is induced,
\begin{equation}
F_{z}^\mathrm{cl} = 0
\label{app_eq:nopinning}
\end{equation}
holds because the average surface lateral force on each fluid particles from the solid is zero.~\cite{Yamaguchi2019,Kusudo2019,Imaizumi2020,Bistafa2021} 
This condition is applicable to the present systems.
By inserting Eqs.~\eqref{eq:fzdiag_usl}, \eqref{eq:fzdiag_usv} and \eqref{app_eq:nopinning} into Eq.~\eqref{app_eq:forcebalance_CL}, 
\begin{equation}
%    \therefore
    \xizcl =  - \usl + \usv = (- \usl) - (-\usv)
\label{app_eq:xizcl_eq_potdif}
\end{equation}
is derived as the analytical expression of $\xizcl$ 
in Eq.~\eqref{eq:xizcl_eq_potdif} in the main text, where the final expression is 
to emphasize 
%appended considering 
that the potential energy densities $\usl$ and $\usv$ are both negative.
%
%
%
%\bibliography{./reference.bib}
\end{document}